\documentclass[prd,amsmath,amssymb,cite,nofootinbib]{revtex4-2} 

\usepackage[utf8]{inputenc}
\usepackage{floatpag}
\usepackage{amsthm}
\usepackage{enumerate}
\usepackage{amssymb}
\usepackage{amsmath}
\usepackage{mathtools}
\usepackage{bbold}
\usepackage{braket}
\usepackage{hyperref}
\usepackage{xcolor}
\definecolor{rosa}{HTML}{E63946}
\definecolor{azul}{HTML}{1D3557}
\hypersetup{
	colorlinks   = true, %Colours links instead of ugly boxes
	urlcolor     = azul, %Colour for external hyperlinks
	linkcolor    = azul, %Colour of internal links
	citecolor   = rosa %Colour of citations
}

\usepackage{graphicx}
\usepackage{epstopdf}
\usepackage{color} % for inkscape
\graphicspath{{./img/}}
\usepackage{caption}
\usepackage{subcaption}
\usepackage[capitalise]{cleveref}

\usepackage{lipsum}  

\begin{document}

\title{Next-to-eikonal corrections to dijet production in Deep Inelastic Scattering in the dilute limit of the Color Glass Condensate}

\author{Pedro Agostini$^{a,b}$, Tolga Altinoluk$^{a}$ and N\'estor Armesto$^{b}$}

\affiliation{
 $^a$ Theoretical Physics Division, National Centre for Nuclear Research, Pasteura 7, Warsaw 02-093, Poland \\
$^b$ Instituto Galego de F\'{\i}sica de Altas Enerx\'{\i}as IGFAE, Universidade de Santiago de Compostela, 15782 Santiago de Compostela, Galicia-Spain
}

\begin{abstract}
	We analyze the effects of next-to-eikonal corrections on dijet production in Deep Inelastic Scattering off nuclear targets in the framework of the Color Glass Condensate. They require the knowledge of correlators of fields in the target beyond those computed in the standard McLerran-Venugopalan model, specifically those between transverse and boost-enhanced components, and of the recoil of the fields. We neglect the latter, while for the former we develop
%	\textcolor{red}{an abelian}
%	
%	\textcolor{red}{\bf P.A.: I think we shouldn't use the word Abelian. Although we are neglecting the non-linear terms of the YM equations, the algebra of the gauge fields is still $\mathrm{su}(N_c)$. I propose using the world linear model but if you find a better one it is welcome.} 
%	
a linear model valid for large nuclei.
	We  considered the unpolarized cross sections for dijet production in the approximation of a homogenous dilute nucleus, obtaining simple analytic expressions for the cross sections at nex-to-eikonal accuracy, valid in the limit of total dijet momentum  and dijet momentum imbalance larger than the saturation scale of the nucleus. We  perform a numerical study of the results at energies of the Electron Ion Collider, finding $\mathcal{O}(10\%)$ effects in the cross sections at large total momentum. We also analyze the azimuthal asymmetries between total momentum and imbalance, finding that non-eikonal corrections induce odd azimuthal harmonics for the situation of jets with equal momentum fractions from the virtual photon, where they are absent in the eikonal approximation. Finally, in the eikonal approximation we have compared the results of our analytic expansion valid in the dilute limit of the target, and the full  Color Glass Condensate results in the McLerran-Venugopalan model and their correlation limit. Our analytic expressions match the correlation limit ones in the region where both should be simultaneously valid and reproduce very well the full  Color Glass Condensate results in its validity region.
\end{abstract}

\maketitle

\section{Introduction}
\label{sec:introduction}
	
%	\textcolor{red}{Particle production at high energy in collisions of a light projectile with heavy nuclei is usually calculated in the framework of the Color Glass Condensate (CGC)~\cite{Gelis:2010nm,Kovchegov:2012mbw}. In this weak coupling but non-perturbative effective field theory, the projectile is treated 
%		as a dilute object whose wave function can be computed order by order in perturbation theory, for which light-cone perturbation theory is often employed~\cite{Brodsky:1997de}. The partons in the projectile then rescatter off the nuclear target which is described as an ensemble of classical fields. The rescatterings are resummed to all orders in the large density of the target which, in the eikonal approximation where terms subleading in energy are neglected , amounts to describing them as Wilson lines at fixed transverse positions.}
	
	Particle production at high energies in collisions of a light projectile with heavy nuclei is typically calculated within the framework of the Color Glass Condensate (CGC)~\cite{Gelis:2010nm,Kovchegov:2012mbw}. This weak-coupling, non-perturbative effective field theory treats the projectile as a dilute object with a wave function computable order-by-order in perturbation theory, often employing light-cone perturbation theory~\cite{Brodsky:1997de}. The projectile partons subsequently rescatter off the nuclear target, modeled as an ensemble of classical fields. These rescatterings are resummed to all orders in the target's high density, which, in the eikonal approximation where terms subleading in energy  are neglected in the calculation of observables, translates to describing them as Wilson lines at fixed transverse positions.	
	
%	\textcolor{red}{
%	The CGC results in non-linear evolution equations for the target ensembles of Wilson lines, that lead to a saturation of the gluon density for those gluons with transverse momenta smaller than the saturation scale of the hadron or nucleus, $Q_s$. Finding evidences of such novel non-linear phenomena in Quantum Chromodynamics and of saturation has been a key point in Deep Inelastic Scattering (DIS) at HERA at DESY, and in proton-proton, proton-nucleus and nucleus-nucleus collisions, and photoproduction in ultraperipheral collisions, at the Relativistic Heavy Ion Collider (RHIC) at BNL and the Large Hadron Collider (LHC) at CERN, see e.g.~\cite{Hentschinski:2022xnd} and references therein. But in spite of claims at HERA~\cite{Ball:2017otu,xFitterDevelopersTeam:2018hym} and RHIC~\cite{BRAHMS:2004xry,PHENIX:2004nzn,STAR:2006dgg,PHENIX:2011puq,STAR:2021fgw}, no conclusive evidence of its existence has been found up to date.}

	The CGC yields non-linear evolution equations for the target ensembles of Wilson lines, leading to gluon density saturation for gluons with transverse momenta smaller than the hadron or nucleus saturation scale, $Q_s$. Unveiling evidence of such novel non-linear Quantum Chromodynamics (QCD) phenomena and of saturation has been a central focus in Deep Inelastic Scattering (DIS) at HERA at DESY, and in proton-proton, proton-nucleus, nucleus-nucleus collisions, and photoproduction in ultraperipheral collisions, at the Relativistic Heavy Ion Collider (RHIC) at BNL and the Large Hadron Collider (LHC) at CERN, see e.g.~\cite{Hentschinski:2022xnd} and references therein. Despite claims at HERA~\cite{Ball:2017otu,xFitterDevelopersTeam:2018hym} and RHIC~\cite{BRAHMS:2004xry,PHENIX:2004nzn,STAR:2006dgg,PHENIX:2011puq,STAR:2021fgw}, no conclusive evidence exists to date.
	
%	\textcolor{red}{
%	In the future, studies will continue at RHIC-II and Runs from 3 on at the LHC, with a dedicated subdetector, FoCal~\cite{ALICE:2020mso}, proposed to be installed in ALICE. Besides, the Electron Ion Collider (EIC)~\cite{Accardi:2012qut,AbdulKhalek:2021gbh}, to be built at BNL and expected to start data taking in the mid 2030's, offers very high luminosities, variable center-of-mass energies in the range $20-140$ GeV/nucleon, and the possibility of multiple nuclear targets and of polarization of both the lepton projectiles and of proton and light ion targets. The search for the saturation regime is one of the pillars of its physics program.}
	
	Future studies will continue at RHIC-II and LHC Runs from 3 onwards, with the dedicated subdetector FoCal~\cite{ALICE:2020mso} proposed for installation in ALICE. Furthermore, the Electron Ion Collider (EIC)~\cite{Accardi:2012qut,AbdulKhalek:2021gbh}, to be built at BNL and expected to begin data acquisition in the mid-2030s, offers exceptionally high luminosities, variable center-of-mass energies ranging from $20-140$ GeV/nucleon, and the possibility of utilizing multiple nuclear targets and both lepton projectile and proton/light ion target polarization. The quest for the saturation regime constitutes one of the pillars of its physics program.
	
%	\textcolor{red}{
%	Concerning the target {\bf gluon} fields, the eikonal approximation amounts to: considering them as localized in the longitudinal direction (called the shockwave approximation); taking only the leading {\bf longitudinal} component component of the field (i.e., that enhanced by the boost), and; neglecting the dynamics of the target. Thus the field has a single component, its longitudinal coordinate  dependence becomes a $\delta$-function and the only remaining coordinate dependence is on the transverse position. For a left moving target and in light-cone coordinates $x^\pm=\left(x^0\pm x^3\right)/\sqrt{2}$, it thus reads
%	\begin{equation}
%		A^\mu_a(x^+,x^-,\mathbf{x})\approx \delta^{\mu -} \delta(x^+) A^-_a(\mathbf{x}),
%	\end{equation}
%	with $a=1,\dots,N_c^2-1$ the color index and boldface letters denoting transverse coordinates.}
	
	The eikonal approximation simplifies the treatment of target gluon fields in three key ways. First, it treats the fields as localized in the longitudinal direction, akin to a "shockwave," neglecting their longitudinal extent. Second, it focuses solely on the leading longitudinal component of the field, the one enhanced by the boost, effectively ignoring other components. Finally, the dynamics of the target itself are disregarded. 
	
	As a result, the target gluon field becomes a single component with a delta-function dependence on the longitudinal coordinate, leaving only the transverse position as relevant variable. In light-cone coordinates $(x^+, x^-)$, where $x^\pm = (x^0 \pm x^3)/\sqrt{2}$ for a left-moving target, this simplified field can be expressed as
	\begin{equation}
		A^\mu_a(x^+,x^-,\mathbf{x}) \approx \delta^{\mu -} \delta(x^+) \alpha_a(\mathbf{x}),
	\end{equation}
	where $a=1,\dots,N_c^2-1$ is the color index, boldface denotes transverse coordinates and the function $\alpha_a$ represents the transverse dependence of the gauge field.
	
%	\textcolor{red}{
%	Studies of non-eikonal corrections in the CGC started around ten years ago relaxing the shockwave approximation~\cite{Altinoluk:2014oxa,Altinoluk:2015gia} by considering a target with a finite longitudinal extent, a central point in studies of medium-induced gluon radiation in the last thirty years, see, e.g., the reviews~\cite{Casalderrey-Solana:2007knd,Mehtar-Tani:2013pia,Blaizot:2015lma}. Later corrections to the other two approximations were computed for scalar, quark and gluon propagators, and introducing quark exchanges with the target~\cite{Altinoluk:2020oyd,Altinoluk:2021lvu,Agostini:2023cvc,Balitsky:2015qba,Balitsky:2016dgz,Chirilli:2018kkw,Chirilli:2021lif}. They are essential in spin physics as spin exchanges are subleading in energy and, therefore, subeikonal~\cite{Kovchegov:2015pbl,Kovchegov:2016weo,Kovchegov:2016zex,Kovchegov:2017jxc,Kovchegov:2018znm,Cougoulic:2019aja,Kovchegov:2021iyc}.}
	
	Studies of non-eikonal corrections in the CGC emerged roughly a decade ago~\cite{Altinoluk:2014oxa,Altinoluk:2015gia}. These studies relaxed the shockwave approximation by considering a target with a finite longitudinal extent, a central focus in medium-induced gluon radiation studies for the past three decades (see, e.g., the reviews~\cite{Casalderrey-Solana:2007knd,Mehtar-Tani:2013pia,Blaizot:2015lma}). Subsequently, corrections to the other two eikonal approximations were calculated for scalar, quark, and gluon propagators, additionally incorporating quark exchanges with the target~\cite{Altinoluk:2020oyd,Altinoluk:2021lvu,Agostini:2023cvc,Balitsky:2015qba,Balitsky:2016dgz,Chirilli:2018kkw,Chirilli:2021lif,Altinoluk:2023qfr,Jalilian-Marian:2017ttv,Jalilian-Marian:2018iui}. These corrections are crucial for spin physics, as spin exchanges are subleading in energy and therefore subeikonal~\cite{Kovchegov:2015pbl,Kovchegov:2016weo,Kovchegov:2016zex,Kovchegov:2017jxc,Kovchegov:2018znm,Cougoulic:2019aja,Cougoulic:2020tbc,Kovchegov:2021iyc,Jalilian-Marian:2019kaf}.
	
%	\textcolor{red}{
%	Numerical studies of the impact of non-eikonal corrections on non-polarized observables have only considered, until now, the effect of the relaxation of the shockwave approximation in single and double inclusive particle production in proton-proton and proton-nucleus collisions~\cite{Altinoluk:2015xuy,Agostini:2019avp,Agostini:2019hkj,Agostini:2022ctk,Agostini:2022oge}. 
%	For the remaining non-eikonal corrections, either the cross sections have not been calculated yet with the required accuracy, or they require additional modelling of the target averages of fields. The generic conclusion of those studies was that the considered non-eikonal effects are sizeable for $\sqrt{s_{NN}}\lesssim 100$ GeV, thus below top RHIC energies. But these are exactly the energies of interest for the EIC, which provides the motivation for examining the size of non-eikonal effects in DIS observables.}

	Until now, numerical studies of the impact of non-eikonal corrections on unpolarized observables have exclusively focused on the relaxation of the shockwave approximation in single and double inclusive particle production in proton-proton and proton-nucleus collisions~\cite{Altinoluk:2015xuy,Agostini:2019avp,Agostini:2019hkj,Agostini:2022ctk,Agostini:2022oge}. For the remaining non-eikonal corrections, either the cross sections have not yet been calculated with sufficient accuracy, or they necessitate further modeling of the target field averages. The general conclusion from these studies suggests that the considered non-eikonal effects are significant for $\sqrt{s_{NN}} \lesssim 100$ GeV, thus below the top RHIC energies. However, these are precisely the energies of interest for the EIC, which provides the motivation for examining the size of non-eikonal effects in DIS observables.
	
%	\textcolor{red}{
%	A large activity has been developed lately on dijet production in DIS due to its potential to disentangle saturation effects~\cite{Dumitru:2018kuw,Mantysaari:2019hkq,Zhao:2021kae,Boussarie:2021ybe,vanHameren:2021sqc}, including the recent computation at next-to-leading order in the coupling constant~\cite{Caucal:2021ent,Taels:2022tza,Caucal:2022ulg,Caucal:2023nci,Caucal:2023fsf}. The relevant cross section including non-eikonal corrections was computed in~\cite{Altinoluk:2022jkk}. But its numerical implementation requires knowledge about the target averages of fields different that $\langle A^- A^-\rangle$ provided by the standard McLerran-Venugopalan (MV) model~\cite{McLerran:1993ni,McLerran:1993ka}. I.e., it requires knowledge about averages that also involve transverse components $A^i$, $\langle A^- A^i \rangle, \langle A^i A^j \rangle$, and also on their $x^-$ dependence. Such information is not available at present beyond the spin contributions computed in~\cite{Cougoulic:2019aja}.}
	
	Recently, dijet production in DIS has garnered significant attention for its potential to reveal saturation effects~\cite{Dumitru:2018kuw,Mantysaari:2019hkq,Zhao:2021kae,Boussarie:2021ybe,Altinoluk:2021ygv,vanHameren:2021sqc}. This includes the recent computation of the cross section at next-to-leading order in the coupling constant~\cite{Caucal:2021ent,Taels:2022tza,Caucal:2022ulg,Caucal:2023nci,Caucal:2023fsf}. Notably, the relevant cross section incorporating non-eikonal corrections was computed in~\cite{Altinoluk:2022jkk}. However, its numerical implementation necessitates knowledge of target field averages beyond the standard $\langle A^- A^-\rangle$ provided by the McLerran-Venugopalan (MV) model~\cite{McLerran:1993ni,McLerran:1993ka}. Specifically, it requires information about averages involving transverse components ${\bf A}^i$, such as $\langle A^- {\bf A}^i \rangle$ and $\langle {\bf A}^i {\bf A}^j \rangle$, and their dependence on the longitudinal coordinate $x^-$. Unfortunately, such information is currently unavailable, except for the spin contributions computed in~\cite{Cougoulic:2020tbc}.
	
%	\textcolor{red}{
%	Additional interest on this observable comes from the fact that in the correlation limit, it offers a link between CGC calculations and transverse momentum distributions (TMD)~\cite{Boussarie:2023izj,Dominguez:2011wm,Kotko:2015ura,Dumitru:2015gaa,Dumitru:2018kuw,Marquet:2016cgx,Altinoluk:2019fui,Altinoluk:2019wyu} which contain three dimensional information about the structure of hadrons and nuclei, and the respective factorization schemes. For two jets $1,2$ with transverse momenta $\mathbf{k}_{1,2}$ and taking $+$ light-cone momentum fractions $z_{1,2}$ from the (virtual) photon, we define the momentum imbalance, ${\bf \Delta} = {\bf k}_1 + {\bf k}_2$, and the relative momentum, ${\bf P} = z_2 {\bf k}_1 - z_1 {\bf k}_2$. The correlation limit corresponds to $|{\bf P}|\gg |{\bf \Delta}|$, and in this limit a relation between the target averages of Wilson lines and the target TMDs at momentum fraction $x=0$ is found.}
	
	Furthermore, dijet production in DIS holds additional interest due to its potential link between CGC calculations and transverse momentum distributions (TMDs) in the so-called "correlation limit"~\cite{Boussarie:2023izj,Dominguez:2011wm,Kotko:2015ura,Dumitru:2015gaa,Dumitru:2018kuw,Marquet:2016cgx,Marquet:2017xwy,Altinoluk:2019fui,Altinoluk:2019wyu}. These TMDs encode three-dimensional information about the hadron and nuclear structure. Consider two jets with transverse momenta $\mathbf{k}_{1,2}$ and light-cone momentum fractions $z_{1,2}$ acquired from the (virtual) photon. We define the momentum imbalance as ${\bf \Delta} = {\bf k}_1 + {\bf k}_2$ and the relative momentum as ${\bf P} = z_2 {\bf k}_1 - z_1 {\bf k}_2$. The correlation limit corresponds to $|{\bf P}| \gg |{\bf \Delta}|$, under which a connection between target averages of Wilson lines and target TMDs at momentum fraction $x=0$ is established.

	This work investigates the impact of non-eikonal effects on dijet production in nuclear DIS. Building upon the framework established in \cite{Altinoluk:2022jkk}, which incorporates next-to-eikonal corrections arising from the transverse component and $x^-$ dependence of the field, we restrict our analysis to the $x^-$ independent case.	To incorporate the transverse component into the target average of field correlators, we develop an extension of the MV model, inspired by the work of \cite{Cougoulic:2020tbc} for polarized targets.
	
	Subsequently, we investigate the dijet production differential cross section in the dilute limit of the CGC\footnote{This approximation corresponds to high energy factorization~\cite{Catani:1990eg,Collins:1991ty} and is also known as the ``glasma graph" approximation in two-particle correlation studies~\cite{Dumitru:2010iy}.}, where partons interact solely through two-gluon exchange. This approximation holds for large values of transverse momenta $|{\bf P}|$ and $|{\bf \Delta}|$, exceeding the saturation scale $Q_s$. Leveraging this approximation, we achieve numerical solutions for the cross section in both the eikonal (see Eqs.~\eqref{eq:xSectionL_Eik} and \eqref{eq:xSectionT_Eik}) and next-to-eikonal (see Eqs.~\eqref{eq:xSectionL_NEik} and \eqref{eq:xSectionT_NEik}) scenarios.
	
	Moreover, we conduct a numerical analysis of the analytical results, focusing on the azimuthal anisotropy expansion of the cross section for $e^-{\rm Au}$ collisions at $\sqrt{s} = 90$ GeV. Our findings reveal significant non-eikonal corrections (up to 10\%) for moderate values of transverse momenta $P_\perp$ and $\Delta_\perp$. Notably, non-eikonal effects induce odd azimuthal harmonics, which are absent in the eikonal approximation.
	
\begin{figure}[h!]
				\centering
				\includegraphics[scale=0.8]{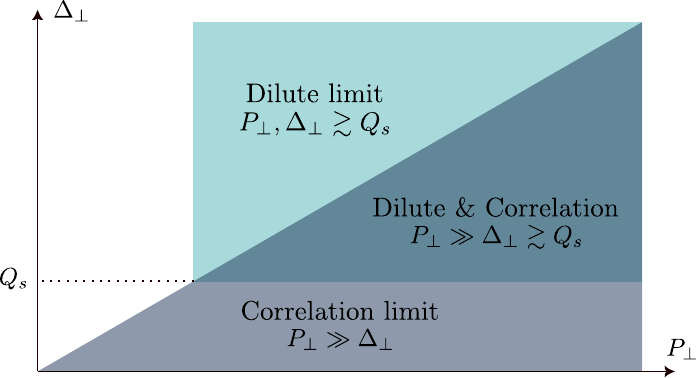}
				\caption{Comparison of our approach with the correlation limit of the CGC. We see that when $\Delta_\perp \ll P_\perp$ we get a good agreement between both approaches as expected.}
				\label{fig:fig1}
			\end{figure}

	To finalize our analysis, we compare the dilute limit with the dense and correlation limit of the CGC. We observe that the eikonal dilute limit of the DIS differential cross section accurately reproduces both the full eikonal CGC result and its correlation limit within the kinematic region $|{\bf P}|,|{\bf \Delta}|\gtrsim Q_s$. The validity of the different approximations is illustrated in Fig.~\ref{fig:fig1}, where $P_\perp=|{\bf P}|,\Delta_\perp= |{\bf \Delta}|$.
	
	The manuscript is organized as follows. In Section~\ref{sec:notation_LCWF} we provide the setup for the calculation. In Section~\ref{sec:field_correlator} we derive the field correlators between the target fields taking into account the $A^+$ and ${\bf A}^i$ components of the background field, absent in the standard MV model. In Section~\ref{sec:DIS_Neik} we review the differential cross section for $q {\bar q}$ production in DIS beyond the eikonal approximation. In Section~\ref{sec:DIS_dilute} we solve the differential cross section in the dilute limit of the CGC. In Section~\ref{sec:numerical_results} we present our numerical results. Section~\ref{sec:conclusion} contains our conclusions. Finally, Appendix~\ref{sec:dense_limit} contains a brief review of the dense limit of the Wilson line averages as well as its correlation limit.
	
\section{Notations and conventions on the nuclear wave function}
\label{sec:notation_LCWF}

	In this section, we employ the Light-Cone Wave Function (LCWF) formalism, detailed in references \cite{Brodsky:1997de, Dumitru:2018vpr, Cougoulic:2020tbc} and their citations, to compute the expectation value of an observable within an ultra-relativistic, large nucleus. Using this formalism, we describe an unpolarized, left-moving nucleus consisting of $A$ nucleons with 3-momenta denoted as $\vec{p}_i \equiv (p_i^-, \mathbf{p}_i)$ via the following wave function:
	\begin{equation}\label{eq:nucleusWF}
		\ket{A} = \int \prod_{i=1}^A \frac{d^3 \vec{p}_i}{2(2 \pi)^3 p_i^-} \psi(\{\vec{p}\}) \ket{\{\vec{p}\}}.
	\end{equation}
	Here, $\{\vec{p}\} \equiv \vec{p}_1, \dots, \vec{p}_A$ represents the 3-momenta of all nucleons, and $\ket{\{\vec{p}\}}$ is the Fock space basis vector for a system with $A$ nucleons. The function $\psi(\{\vec{p}\}) = \braket{\{\vec{p}\} | A}$ denotes the amplitude associated with a specific momentum configuration. We normalize the nuclear wave function as $\braket{A | A} = 1$, ensuring that the light-cone amplitude also satisfies the normalization condition:
	\begin{equation}\label{key}
		\int \prod_{i=1}^A \frac{d^3 \vec{p}_i}{2(2 \pi)^3 p_i^-} |\psi(\{\vec{p}\})|^2=1.
	\end{equation}
	
	Furthermore, the single-particle state of a nucleon with momentum $\vec{p}$ fulfills
	\begin{equation}\label{key}
		\braket{\vec{p}' | \vec{p}}= 2 p^- (2 \pi)^3 \delta^{(3)}(\vec{p}-\vec{p}').
	\end{equation}
	
	Referring to~\cref{eq:nucleusWF}, and considering an observable characterized by the operator $\hat{\mathcal{O}}$, its expectation value within the nuclear ensemble can be expressed as follows:	
	\begin{align}\label{eq:aux4}
		\langle \hat{O} \rangle_A \equiv \bra{A} \hat{\mathcal{O}} \ket{A} &=
		\int \prod_{i=1}^A \frac{d^3 \vec{P}_i}{2(2 \pi)^3 \left( P_i^- + \frac{q_i^-}{2}\right)} \frac{d^3 \vec{q}_i}{2(2 \pi)^3 \left( P_i^- - \frac{q_i^-}{2}\right)} 
		\psi^*\left(\Big\{\vec{P} - \frac{\vec{q}}{2}\Big\}\right) \psi\left(\Big\{\vec{P} + \frac{\vec{q}}{2}\Big\}\right) \nonumber \\
		& \times
		\Big\langle \Big\{\vec{P} - \frac{\vec{q}}{2}\Big\} \Big| \hat{\mathcal{O}} \Big| \Big\{\vec{P} + \frac{\vec{q}}{2}\Big\} \Big\rangle.
	\end{align}
	
	For convenience, we can express~\cref{eq:aux4} in terms of the Wigner distribution of the quantum system by rephrasing it as follows:	
	\begin{align}\label{eq:avg_nucl}
		\langle \hat{O} \rangle_A =
		\prod_{i=1}^A \int \frac{d^3 \vec{P}_i}{(2 \pi)^3} \int d^3 \vec{B}_i W(\{\vec{P},\vec{B}\}) O(\{\vec{P},\vec{B}\}),
	\end{align}	
	where
	\begin{equation}\label{key}
		W(\{\vec{P},\vec{B}\}) = \int \prod_{i=1}^A \frac{d^3 \vec{q}_i}{2(2 \pi)^3} \frac{P_i^-}{\left(P_i^- + \frac{q_i^-}{2}\right) \left(P_i^- - \frac{q_i^-}{2}\right)} e^{-i \vec{q}_i \cdot \vec{B}_i}
		\psi^*\left(\Big\{\vec{P} - \frac{\vec{q}}{2}\Big\}\right) \psi\left(\Big\{\vec{P} + \frac{\vec{q}}{2}\Big\}\right)
	\end{equation}
	represents the multi-particle Wigner distribution of the nuclear ensemble, which provides information on the spatial and momentum distribution of the nucleons within the nucleus. Furthermore, we have defined
	\begin{equation}\label{eq:aux1}
		O(\{\vec{P},\vec{B}\}) = \int \prod_{i=1}^A \frac{d^3 \vec{q}_i}{(2 \pi)^3 2 P_i^-} e^{i \vec{q}_i \cdot \vec{B}_i}
		\Big\langle \Big\{ \vec{P} - \frac{\vec{q}}{2} \Big\} \Big| \hat{\mathcal{O}} \Big| \Big\{ \vec{P} + \frac{\vec{q}}{2} \Big\} \Big\rangle,
	\end{equation}
	which accounts for the expectation value of the observable at a given configuration of the nucleons inside the nucleus.
	
	While~\cref{eq:avg_nucl} is entirely general and model-independent, in practical applications the Wigner distribution is notoriously difficult to calculate directly and often requires specification through a particular theoretical framework. Given the vast number of nucleons comprising the nucleus, correlations between them can often be disregarded as they scale as $1/\sqrt{A}$. This allows us to factorize the amplitude for the multi-particle system as a product of single-particle amplitudes, leading to a more tractable expression for the Wigner distribution:
	\begin{equation}
		\label{eq:Wigner_fact}
		W(\{\vec{P},\vec{B}\}) = \prod_{i=1}^A W(\vec{P}_i,\vec{B}_i).
	\end{equation}
	Here, it is important to note that we use the same notation, $W$, to represent both the multi-particle and single-particle Wigner distributions.
	
	Ultimately, the single-particle Wigner function can be described using various models. Here, we adopt a Glauber-like model where the momentum and spatial distributions are factorized:
	\begin{equation}
		\label{eq:Wigner_model}
		W(\vec{P},\vec{B}) = \zeta(\vec{P})\, \frac{\rho_A(\vec{B})}{A},
	\end{equation}
	where $\rho_A(\vec{x})$ represents the normalized nucleon density, such that $\int d^3 \vec{x} \rho_A(\vec{x}) = A$. Additionally, $\zeta(\vec{P}) \equiv |\phi(\vec{P})|^2/2P^-$ accounts for the momentum distribution of the nucleons within the nucleus.
	
	Following the conventions of the McLerran-Venugopalan model as detailed in~\cite{Cougoulic:2020tbc}, we assume that each nucleon possesses a common, large longitudinal momentum denoted as $P^-_{\text{N}}=P^-_A/A$, where $\vec{P}_A$ corresponds to the 3-momentum of the nucleus. Consequently, for a nucleus with negligible transverse momentum, we can write\footnote{In principle, for a nucleus with negligible transverse momentum, the individual transverse momenta of the nucleons do not necessarily have to be zero. However, in the context of this analysis, the expectation value will solely rely on the average transverse momentum of the nucleons, which will ultimately approach zero. To streamline our analysis, we make the simplifying assumption that the nucleons carry no transverse momentum.}
	\begin{equation}
		\label{eq:MV_model}
		\zeta(\vec{P})= (2 \pi)^3 \delta(P^--P^-_{\rm N}) \delta^{(2)}({\bf P}).
	\end{equation}
	This essentially implies that all nucleons share the same large longitudinal momentum ($P^-_{\rm N}$), while their transverse momenta are distributed around zero.
	
	By substituting~\cref{eq:Wigner_fact,eq:Wigner_model,eq:MV_model} into~\cref{eq:avg_nucl}, we arrive at a simplified expression for the expectation value:
	\begin{align}
		\label{eq:avg_nucl2}
		\langle \hat{O} \rangle_A =
		\prod_{i=1}^A \int d^3 \vec{B}_i \frac{\rho_A(\vec{B}_i)}{A} O\left(\{\vec{P}=(P^-_{\rm N}, 0), \vec{B}\}\right).
	\end{align}
	This result highlights that the expectation value only depends on the integral over the nuclear density $\rho_A(\vec{B}_i)$ and the observable evaluated at the common longitudinal momentum $P^-_{\rm N}$ and zero transverse momentum.
	
	In the CGC framework, the nuclear wave function is characterized by sources represented as large-$x$ recoilless valence partons, while the small-$x$ gluons are determined by the solutions of the classical Yang-Mills equations associated with these sources. Consequently, the expectation values of observables are computed exclusively through the degrees of freedom of the valence partons. Under the assumption that each nucleon is comprised of $N_c$ quarks\footnote{While a complete non-eikonal treatment would necessitate incorporating ``valence" gluons and antiquarks into the nuclear wave function, as demonstrated in~\cite{Cougoulic:2020tbc}, it is crucial to emphasize that in the present study eikonal corrections solely focus on the classical gluon field generated by the valence quarks.} such that the nucleons are singlets, the single nucleon state is written as follows~\cite{Dumitru:2018vpr}:
	\begin{equation}\label{key}
		\ket{\vec{p}} = \frac{1}{\sqrt{N_c!}} \int \prod_{i=1}^{N_c} \frac{d^3 \vec{k}_i}{2(2 \pi)^3 k_i^-} \Psi(\{\vec{k}\}) \epsilon_{\alpha_1,\dots,\alpha_{N_c}} \ket{\{\vec{k},\alpha\}}
		2(2 \pi)^3 p^- \delta^{(3)}\left(
		\vec{p} - \sum_{n=1}^{N_c} \vec{k}_{n}
		\right),
	\end{equation}
	where the basis for the Fock Space of a system with $N_c$ quarks is denoted as $\ket{\{\vec{k},\alpha\}} = \ket{\vec{k}_1,\alpha_1} \otimes \cdots \otimes \ket{\vec{k}_{N_c},\alpha_{N_c}}$. Here, $\ket{\vec{k},\alpha}$ represents a single quark state characterized by momentum $\vec{k}$ and color $\alpha$, while $\Psi(\{\vec{k}\})$ represents the amplitude associated with this particular configuration. The Levi-Civita tensor, denoted as $\epsilon_{\alpha_1,\dots,\alpha_{N_c}}$, ensures that the state is a singlet. It is important to mention that, since this study assumes an unpolarized nucleus, we omit the helicity quantum numbers. The single quark states are defined by $\ket{\vec{k},\alpha} = \hat{b}^\dagger_{\alpha}(\vec{k}) \ket{0}$, with the creation and annihilation operators adhering to the following normalization:
	\begin{equation}\label{eq:commutation_relations}
		\{ \hat{b}_{\alpha'}(\vec{k}'), \hat{b}^\dagger_{\alpha}(\vec{k}) \} = \delta_{\alpha \alpha'} 2 k^-(2\pi)^3 \delta^{(3)}(\vec{k}-\vec{k}'), \qquad \{ \hat{b}^\dagger_{\alpha'}(\vec{k}'), \hat{b}^\dagger_{\alpha}(\vec{k}) \} = \{ \hat{b}_{\alpha'}(\vec{k}'), \hat{b}_{\alpha}(\vec{k}) \} = 0.
	\end{equation}
	
	The state of $A$ color-neutral and independent nucleons composed by $N_c$ quarks is represented as
	\begin{equation}\label{key}
		\ket{\{\vec{p}\}} = \prod_{i=1}^A \left[\frac{1}{\sqrt{N_c!}} \int \prod_{j=1}^{N_c} \frac{d^3 \vec{k}_{ij}}{2(2 \pi)^3 k_{ij}^-} \epsilon_{\{\alpha_i\}} \Psi(\{\vec{k}_i\}) \ket{\{\vec{k}_i,\alpha_i\}} 
		16 \pi^3 p^- \delta^{(3)}\left(
		\vec{p} - \sum_{n=1}^{N_c} \vec{k}_{in}
		\right)
		\right] ,
	\end{equation}
	where we use the notation $\{\alpha_i\} \equiv \alpha_{i1}, \dots, \alpha_{i N_c}$.
	
	We begin by assuming that quarks within the nucleon wave function behave independently, separated by distances roughly equivalent to the nucleon's size ($r_{\rm N} \sim \Lambda_{\rm QCD}^{-1}$, where $\Lambda_{\rm QCD}$ denotes the QCD non-perturbative scale). This allows us to make similar assumptions as done at the nucleon level and introduce the quark distribution, $\rho_{N_c}(\vec{x})$, normalized such that $\int d^3 \vec{x} \rho_{N_c}(\vec{x}) = N_c$. All in all, we arrive at the following expression:
	\begin{equation}\label{eq:matrix_nucleon}
		\bra{\{\vec{p}'\}} \hat{\mathcal{O}} \ket{\{\vec{p}\}} = 
		\frac{1}{A!}\Big( \braket{\vec{p}_1' | \vec{p}_1} \cdots \braket{\vec{p}_A' | \vec{p}_A}  + {\rm permutations} \Big) \prod_{i=1}^A \prod_{j=1}^N \int d^3 \vec{b}_{ij} \frac{\rho_N(\vec{b}_{ij}-\vec{B}_i)}{N} 
		\langle \hat{\mathcal{O}} \rangle_q (\{ \vec{b} \}),
	\end{equation}
	where $\langle \hat{\mathcal{O}} \rangle_q (\{ \vec{b} \})$ is the expectation value of the operator at a given quark configuration, defined as follows:
	\begin{equation}\label{eq:obs_moment}
		\langle \hat{\mathcal{O}} \rangle_q (\{ \vec{b} \}) = \prod_{i=1}^A \left[\frac{1}{N_c!} \left(\prod_{j=1}^{N_c} \int \frac{d^3 \vec{q}_{ij}}{2(2\pi)^3 P_q^-} e^{i \vec{q}_{ij} \cdot \vec{b}_{ij}}\right) \epsilon_{\{\alpha_i\}} \epsilon_{\{\bar{\alpha}_i\}}\right] 
		\Big\langle \Big\{P_q^- -\frac{q^-}{2},-\frac{\bf q}{2},\bar{\alpha} \Big\} \Big| \hat{\mathcal{O}} \Big| \Big\{P_q^- +\frac{q^-}{2},\frac{\bf q}{2},\alpha \Big\} \Big\rangle.
	\end{equation}
	Here, $P_q^- \equiv P_{A}^-/AN_c$ defines the longitudinal momentum of quarks within the nucleus. Since $P_A^- = M_A e^{\omega}$ (where $M_A$ and $\omega$ represent the nucleus's mass and rapidity, respectively), $P_q^-$ becomes the dominant momentum scale when the nucleus is highly boosted.
	
	Finally, by substituting~\cref{eq:matrix_nucleon} into~\cref{eq:aux1}, we can express the expectation value of the observable $\hat{\mathcal{O}}$ in the nuclear ensemble, as presented in~\cref{eq:avg_nucl2}, as follows:
	\begin{equation}\label{}
		\langle \hat{O} \rangle_A =
		\prod_{i=1}^{A}\prod_{j=1}^{N_c}
		\int d^3 \vec{B}_i \frac{\rho_A(\vec{B}_i)}{A} \int d^3 \vec{b}_{ij} \frac{\rho_{N_c}(\vec{b}_{ij}-\vec{B}_i)}{N_c} \langle \hat{\mathcal{O}} \rangle_q (\{ \vec{b} \}).
	\end{equation}
	
	We can further approximate by recognizing that $\rho_{N_c}(\vec{b}-\vec{B})$ is predominantly centered around $\vec{B} = \vec{b}$ and confined within the nucleon's size. Meanwhile, $\rho_A(\vec{B})$ extends across the entire nuclear volume, allowing us to treat it as nearly constant over the scale of a single nucleon. With this, we can write
	\begin{equation}\label{eq:targ_avg}
		\langle \hat{O} \rangle_A =
		\prod_{i=1}^{A}\prod_{j=1}^{N_c}
		\int d^3 \vec{b}_{ij}  \frac{\rho_A(\vec{b}_{ij})}{A} \langle \hat{\mathcal{O}} \rangle_q (\{ \vec{b} \}).
	\end{equation}
	
\section{Color source and field correlator}	
\label{sec:mean_value_current}
	
	We turn our attention to computing the expectation value of the color current density correlator within the nuclear state discussed previously. As a first step, we derive the expectation value of the current operator for illustrative purposes. We then proceed to calculate the correlator itself.
	
\subsection{The color current density operator}
	
	The color current density operator, at light-cone time $x^- = 0$, is given in terms of the fermion fields by the following expression:
	\begin{equation}\label{eq:aux2}
		\hat{J}_a^\mu(x^- = 0, \vec{x}) = g \bar{\psi}_\alpha (\vec{x}) \gamma^\mu  t^a_{\alpha \beta} \psi_\beta(\vec{x}).
	\end{equation}
	In this equation, $g$ represents the QCD coupling constant, $t^{a}$ denotes the SU($N_c$) generators in the fundamental representation, $\gamma^\mu$ corresponds to the gamma matrices, and the fermion fields with color $\alpha$ are given by
	\begin{align}\label{eq:aux3}
		\psi_\alpha(\vec{x}) = \int_{\vec{p}} \hat{b}_{\alpha, \sigma}(\vec{p}) \frac{u_\sigma(\vec{p})}{\sqrt{2 p^-}} e^{-i \vec{p} \cdot \vec{x}}, \qquad 
		\bar{\psi}_\alpha(\vec{x}) = \int_{\vec{p}} \hat{b}^\dagger_{\alpha, \sigma}(\vec{p}) \frac{\bar{u}_\sigma(\vec{p})}{\sqrt{2 p^-}} e^{i \vec{p} \cdot \vec{x}}.
	\end{align}
	Here, we have introduced the notation
	\begin{equation}\label{key}
		\int_{\vec{p}} \equiv \int \frac{d^3 \vec{p}}{(2\pi)^3 \sqrt{2 p^-}}.
	\end{equation}
	We exclude the antiquark component of the field due to the exclusively quark composition of the nucleus, as the antiquark component holds no relevance to our present analysis. While we introduce the helicity index $\sigma$ temporarily, it is to be noted that the nucleus is unpolarized, necessitating an averaging over helicity at the end of the calculation.
	
	Therefore, by substituting \cref{eq:aux3} into \cref{eq:aux2}, we can express the color current density as follows:
	\begin{equation}\label{eq:current_operator}
		\hat{J}_a^\mu(\vec{x}) = g t_{\alpha \beta}^a \int_{\vec{p},\vec{p}'} e^{-i \vec{x} \cdot (\vec{p}-\vec{{p}}')} 
		\Gamma^\mu_{\sigma {\sigma}'}(\vec{p},\vec{{p}}')
		\
		\hat{b}_{\alpha, {\sigma}'}^\dagger (\vec{{p}}') \hat{b}_{\beta, \sigma}(\vec{p}),
	\end{equation}
	where
	\begin{align}
		\Gamma^\mu_{\sigma, {\sigma}'}(\vec{p},\vec{{p}}') = \frac{\bar{u}_{{\sigma}'}(\vec{p}')}{\sqrt{2 p'^-}} \gamma^\mu \frac{u_\sigma(\vec{p})}{\sqrt{2p^-}}.
	\end{align}
	To compute this object, we can utilize the Brodsky-Lepage spinors for left-moving massless fermions~\cite{Brodsky:1989pv,Cougoulic:2020tbc}:
	\begin{align}
		u_\sigma(\vec{p}) = \frac{1}{\sqrt{\sqrt{2}p^-}}
		\left(\sqrt{2}p^- + \gamma^0 \gamma^i {\bf p}^i \right) \chi_\sigma,
	\end{align}
	where $\chi_+ = 1/\sqrt{2}(1,0,-1,0)^T$ and $\chi_- = 1/\sqrt{2}(0,1,0,1)^T$, such that
	\begin{subequations}\label{eq:gamma}
		\begin{align}
			\Gamma^-_{\sigma {\sigma}'}(\vec{p},\vec{{p}}') &= \delta_{\sigma \sigma'},
			\\
			\Gamma^+_{\sigma {\sigma}'}(\vec{p},\vec{{p}}')
			&=
			\delta_{\sigma \sigma'} \frac{{\bf p}^i}{p^-} 
			\frac{{\bf p}'^i - i \sigma \epsilon^{ij} {\bf p}'^j}{2 p'^-},
			\\
			\Gamma^i_{\sigma {\sigma}'}(\vec{p},\vec{{p}}')
			&=
			\delta_{\sigma \sigma'} \left(
			\frac{{\bf p}'^i-i \sigma \epsilon^{ij} {\bf p}'^j}{2 p'^-} +
			\frac{{\bf p}^i+i \sigma \epsilon^{ij} {\bf p}^j}{2 p^-}
			\right).
			\label{eq:Gammai}
		\end{align}
	\end{subequations}
	We observe that the transverse component of the current, represented by the helicity tensor defined in~\cref{eq:Gammai}, was previously investigated in \cite{Cougoulic:2020tbc} for the specific case of a polarized target within the study of helicity-dependent processes.
	
	To calculate the expectation value of the density current presented in~\cref{eq:current_operator}, we leverage the following identities derived from the commutation relations in~\cref{eq:commutation_relations}. Within an $N$-particle Fock basis characterized by momentum and color quantum numbers denoted as $s \equiv \vec{k},\alpha$, the following identities apply:
	\begin{subequations}
		\begin{align}
			\hat{b}_{\beta}(\vec{p}) \ket{ s_1, \dots, s_N } &= \sum_{i=1}^{N} (-1)^{i+1} \delta_{\beta \alpha_i} 16 \pi^3 p^- \delta^{(3)}(\vec{p} - \vec{k}_i)
			\ket{s_1, \dots, s_{i-1},s_{i+1},\dots,s_{N}},
			\\
			\hat{b}_{\alpha}^\dagger (\vec{{p}}^{\,\prime}) \hat{b}_{\beta}(\vec{p}) \ket{ s_1, \dots, s_N } &= \sum_{i=1}^{N} \delta_{\beta \alpha_i} 16 \pi^3 p^- \delta^{(3)}(\vec{p} - \vec{k}_i)
			\ket{s_1, \dots ,s_{N} \ \slash \ s_i = \vec{p}^{\,\prime},\alpha }.
		\end{align}
	\end{subequations}
	Consequently, the matrix element of the operator $\hat{b}_{\alpha}^\dagger (\vec{{p}}^{\,\prime}) \hat{b}_{\beta}(\vec{p})$ can be expressed as
	\begin{align}\label{eq:1body_matrix_element}
		\bra{s_1^\prime, \cdots, s_N^\prime} \hat{b}_{\alpha}^\dagger (\vec{{p}}^{\,\prime}) \hat{b}_{\beta}(\vec{p}) \ket{ s_1, \dots, s_N } &= \sum_{i=1}^{N} \delta_{\beta \alpha} 16 \pi^3 p^- \delta^{(3)}(\vec{p} - \vec{k}_i)
		16 \pi^3 p^{\prime -} \delta^{(3)}(\vec{p}^{\,\prime} - \vec{k}^\prime_i)
		\nonumber \\ & \times
		\prod_{j \ne i}^N \delta_{\alpha_j \alpha^\prime_j} 16 \pi^3 k_j^- \delta^{(3)}(\vec{k}_j - \vec{k}^\prime_j) + \cdots,
	\end{align}
	where we have retained only the diagonal term, which is the component that dominates when the particles are independent.
	
	Thus, by using \cref{eq:1body_matrix_element,eq:current_operator,eq:obs_moment} and performing some algebraic manipulations, we arrive at the following expression:
	\begin{align}\label{}
		\langle \hat{J}_a^\mu(\vec{x}) \rangle_q (\{ \vec{b} \}) &= 
		\frac{g {\rm Tr}[t^a]}{N_c}
		\sum_{i=1}^{A} \sum_{j=1}^{N_c}
		\int \frac{d^3 \vec{q}_{ij}}{(2\pi)^3} e^{-i \vec{q}_{ij} \cdot (\vec{x} - \vec{b}_{ij})}
		\frac{1}{2}  \Gamma_{\sigma \sigma}^\mu \left({P}_q^- + \frac{q_{ij}^-}{2}, \frac{{\bf q}_{ij}}{2}; {P}_q^- - \frac{q_{ij}^-}{2}, -\frac{{\bf q}_{ij}}{2} \right)
		\sqrt{1 - \left(  \frac{q_{ij}^-}{2P_q^-} \right)^2}
		\nonumber \\ & \approx \frac{g {\rm Tr}[t^a]}{N_c}
		\sum_{i=1}^{A} \sum_{j=1}^{N_c}
		\int \frac{d^3 \vec{q}}{(2\pi)^3} e^{-i \vec{q} \cdot (\vec{x} - \vec{b}_{ij})}
		\frac{1}{2} \Gamma_{\sigma \sigma}^\mu \left({P}_q^-, \frac{{\bf q}}{2}; {P}_q^-, -\frac{{\bf q}}{2} \right).
	\end{align}
	Here, we have taken an average over helicity by performing the trace over $\sigma$ and dividing by 2. Furthermore, we have neglected terms proportional to $q_{ij}^-$ because they scale as $q_{ij}^- \sim 1/A$ after Fourier transformation, therefore they are subdominant in the large-$A$ limit. Ultimately, the expectation value of the color current density is proportionate to ${\rm Tr}[t^a]=0$ and, hence, it vanishes, in line with the expectation for color-neutral nucleons.
	
\subsection{Color current density correlator}
\label{subsec:current_correlator}
	
	We proceed to calculate the color current density correlator within the nucleus. The operator for the helicity-averaged density correlator is expressed as
	\begin{align}
	\label{eq:aux11}
		\hat{J}_a^\mu(\vec{x}) \hat{J}_b^\nu(\vec{y}) = g^2 t_{\alpha_1 \beta_1}^a t_{\alpha_2 \beta_2}^b \int_{\vec{p}_1,\vec{p}_1^{\,\prime},\vec{p}_2,\vec{p}_2^{\,\prime}}
		& e^{-i \vec{x} \cdot (\vec{p}_1-\vec{p}_1^{\,\prime})-i \vec{y} \cdot (\vec{p}_2-\vec{p}_2^{\,\prime})} 
		\frac{1}{2}
		\Gamma^\mu_{\sigma_1 \sigma_2}(\vec{p}_1,\vec{p}_1^{\,\prime}) \Gamma^\nu_{\sigma_2 \sigma_1}(\vec{p}_2,\vec{p}_2^{\,\prime})
		\nonumber \\ & \times 
		\hat{b}_{\alpha_1}^\dagger (\vec{p}_1^{\,\prime}) \hat{b}_{\beta_1}(\vec{p}_1)
		\hat{b}_{\alpha_2}^\dagger (\vec{p}_2^{\,\prime}) \hat{b}_{\beta_2}(\vec{p}_2).
	\end{align}
	
	For practical applications, we focus on the regime where $|\vec{x} - \vec{y}| \ll \Lambda_{\rm QCD}^{-1}$, which permits the application of weak coupling techniques. Based on the assumption that valence quarks have a typical separation of the order of $\Lambda_{\rm QCD}^{-1}$, both currents probe the same quark within the perturbative regime. From an operator perspective, this implies that the one-body component in~\cref{eq:aux11} dominates over the two-body component. Consequently, we can make the following approximation:
	\begin{align}
		\hat{J}_a^\mu(\vec{x}) \hat{J}_b^\nu(\vec{y}) &\approx g^2 t_{\alpha_1 \beta_1}^a t_{\alpha_2 \beta_2}^b \int_{\vec{p}_1,\vec{p}_1^{\,\prime},\vec{p}_2,\vec{p}_2^{\,\prime}} e^{-i \vec{x} \cdot (\vec{p}_1-\vec{p}_1^{\,\prime})-i \vec{y} \cdot (\vec{p}_2-\vec{p}_2^{\,\prime})} 
		\frac{1}{2}
		\Gamma^\mu_{\sigma_1 \sigma_2}(\vec{p}_1,\vec{p}_1^{\,\prime}) \Gamma^\nu_{\sigma_2 \sigma_1}(\vec{p}_2,\vec{p}_2^{\,\prime})
		\nonumber \\ & \times 
		\delta_{\alpha_2 \beta_1} 16 \pi^3 p_1^- \delta^{(3)}(\vec{p}_1 - \vec{p}_2^{\,\prime}) 
		\hat{b}_{\alpha_1}^\dagger (\vec{p}_1^{\,\prime}) \hat{b}_{\beta_2}(\vec{p}_2)
		\nonumber \\ &=
		g^2 t_{\alpha_1 \beta_1}^a t_{\beta_1 \beta_2}^b \int_{\vec{p}_1,\vec{p}_1^{\,\prime},\vec{p}_2} e^{-i \vec{x} \cdot (\vec{p}_1-\vec{p}_1^{\,\prime})-i \vec{y} \cdot (\vec{p}_2-\vec{p}_1)} \sqrt{2 p_1^-} 
		\frac{1}{2}
		\Gamma^\mu_{\sigma_1 \sigma_2}(\vec{p}_1,\vec{p}_1^{\,\prime}) \Gamma^\nu_{\sigma_2 \sigma_1}(\vec{p}_2,\vec{p}_1)
		\hat{b}_{\alpha_1}^\dagger (\vec{p}_1^{\,\prime}) \hat{b}_{\beta_2}(\vec{p}_2).
	\end{align}
	
	By once more employing~\cref{eq:1body_matrix_element,eq:obs_moment}, in conjunction with the aforementioned approximations, we can derive the following expression:
	\begin{align}\label{}
		\Bigg\langle 
		\hat{J}_a^\mu\left(\vec{B} + \frac{\vec{R}}{2}\right) \hat{J}_b^\nu\left(\vec{B} - \frac{\vec{R}}{2}\right)
		\Bigg\rangle_q (\{\vec{b}\})
		&=  
		\frac{g^2 {\rm Tr}[t^a t^b]}{N_c}
		\sum_{i=1}^{A} \sum_{j=1}^{N_c}
		\int \frac{d^3 \vec{q}}{(2\pi)^3} \frac{d^3 \vec{p}}{(2\pi)^3} e^{-i \vec{q} \cdot (\vec{B} - \vec{b}_{ij}) -i  \vec{p} \cdot \vec{R}} 
		\nonumber \\ & \hskip0cm \times
		\frac{1}{2}\,
		\Gamma^\mu_{\sigma_1 \sigma_2}\left(P_q^- + p^-, {\bf p}; P_q^-,  -\frac{\textbf{q}}{2} \right)
		\Gamma^\nu_{\sigma_2 \sigma_1}\left(P_q^-, \frac{\textbf{q}}{2}; P_q^- + p^-, {\bf p}\right).
	\end{align}

	The dependence of the integrand on $p^-$ results in a non-trivial correlation function, $F(R^+; P_q^-)$, after the transformation. This function introduces a correlation length of $\lambda^+ \sim 1/P_q^-$. Notably, the longitudinal correlations become localized in the eikonal limit, meaning $F(R^+; P_q^- \to \infty) \sim \delta(R^+)$. For practical applications, where we integrate $R^+$ over the entire nuclear width, the contributions from these longitudinal correlations scale as $\lambda^+/L^+ \sim A^{-1/3}$, where $L^+$ represents the light-cone longitudinal width of the medium. Consequently, these contributions become subdominant for large nuclei. For this reason, we do not take into account corrections due to longitudinal particle correlations\footnote{While neglecting the longitudinal correlations based on their subdominant scaling, we acknowledge that their non-eikonal effects could be explicitly incorporated in a more practical way by modifying the $\delta(R^+)$ function. As explored in references \cite{Altinoluk:2015xuy,Agostini:2019avp}, this can be achieved by substituting $\delta(R^+)$ with $\Theta(\lambda^+ - |R^+|)/2 \lambda^+$, where $\Theta$ represents the Heaviside function.
	} in the present manuscript and we drop dependence of the integrand on $p^-$:
	\begin{align}\label{eq:aux5}
		\Bigg\langle 
		\hat{J}_a^\mu\left(\vec{B} + \frac{\vec{R}}{2}\right) \hat{J}_b^\nu\left(\vec{B} - \frac{\vec{R}}{2}\right)
		\Bigg\rangle_q (\{\vec{b}\}) &=  
		\frac{g^2 \delta^{ab}}{N_c}
		\sum_{i=1}^{A} \sum_{j=1}^{N_c}
		\int \frac{d^3 \vec{q}}{(2\pi)^3} \frac{d^3 \vec{p}}{(2\pi)^3} e^{-i \vec{q} \cdot (\vec{B} - \vec{b}_{ij}) -i  \vec{p} \cdot \vec{R}} 
		\nonumber \\ & \hskip0cm \times
		\frac{1}{2}
		\Gamma^\mu_{\sigma_1 \sigma_2}\left(P_q^-, {\bf p}; P_q^-,  -\frac{\textbf{q}}{2} \right)
		\Gamma^\nu_{\sigma_2 \sigma_1}\left(P_q^-, \frac{\textbf{q}}{2}; P_q^-, {\bf p}\right),
	\end{align}
	where we have written ${\rm Tr}[t^a t^b] = \delta^{ab}/2$.
	
	Finally, by substituting \cref{eq:aux5} into \cref{eq:avg_nucl2}, we arrive at the expression for the color current density correlator in a large nucleus\footnote{We use the following definitions for the transverse integrals:
		\begin{align}
			\int_{\bf x} \equiv \int d^2 {\bf x}, \qquad \int_{\bf p} \equiv \frac{d^2 {\bf p}}{(2 \pi)^2}.
		\end{align}
	}:
	\begin{align}\label{eq:current_correlator}
		\Bigg\langle \hat{J}_a^\mu\left(\vec{b} + \frac{\vec{r}}{2}\right) \hat{J}_b^\nu\left(\vec{b} - \frac{\vec{r}}{2}\right) \Bigg\rangle_A &= \delta^{ab} \delta(r^+)
		\int_{\textbf{B}} \mu^2(b^+, \textbf{b} - \textbf{B})
		\int_{{\bf q}, {\bf p}} e^{i \textbf{q} \cdot \textbf{B} + i   \textbf{p} \cdot \textbf{r}} 
		\nonumber \\ & \times 
		\frac{1}{2}
		\Gamma^\mu_{\sigma_1 \sigma_2}\left(P_q^-, {\bf p}; P_q^-,  -\frac{\textbf{q}}{2} \right)
		\Gamma^\nu_{\sigma_2 \sigma_1}\left(P_q^-, \frac{\textbf{q}}{2}; P_q^-, {\bf p}\right),
	\end{align}
	where we have defined
	\begin{align}
		\mu^2(\vec{x}) = \frac{g^2}{2} \rho_A(\vec{x}),
	\end{align}
	which can be interpreted as the mean color charge density (squared) per unit volume.
	
\subsection{Color field correlation}
\label{sec:field_correlator}
	
	Within the CGC framework, observables depend on the correlators of the color field generated by nuclear current sources. When considering a highly boosted, large nucleus, the resultant color field is strong and effectively classical. Consequently, it can be expressed in terms of the color sources via solutions to the classical Yang-Mills (YM) equations. However, solving these equations for a strong, non-eikonal field presents significant analytical and numerical challenges due to their inherent non-linearity. 
	
	To address this complexity, we adopt the dilute limit of the CGC. This allows us to treat the field as weak yet classical, enabling us to get rid of the subleading non-linear terms in the YM equations and enabling analytical solutions even for non-eikonal configurations. Furthermore, we restrict ourselves to light-cone time-independent fields. Under these assumptions, the classical YM equation in covariant gauge ($\partial_\mu A^\mu = 0$) for a time-independent field takes the following form:
	\begin{align}
		\partial^2 \hat{A}_a^\mu(\vec{x})= - \partial_\perp^2 \hat{A}_a^\mu(\vec{x})  = \hat{J}_a^\mu(\vec{x}),
	\end{align}
	with the solution
	\begin{align}
		\hat{A}_a^\mu(\vec{x}) = \int_{\bf z} \int_{\bf q} e^{i{\bf q} \cdot ({\bf x}-{\bf z})}
		\frac{\hat{J}_a^\mu(x^+,{\bf z})}{{\bf q}^2}.
	\end{align}
	
	Consequently, the correlation of the color field within the nuclear ensemble can be written as
	\begin{align}\label{eq:aux6}
		\big\langle \hat{A}_a^\mu(\vec{x}) \hat{A}_b^\nu(\vec{y}) \big\rangle_A
		=
		\int_{ {\bf z}_1, {\bf z}_2 } \int_{ {\bf q}_1, {\bf q}_2 }  e^{ i{\bf q}_1 \cdot ({\bf x}-{\bf z}_1) + i{\bf q}_2 \cdot ({\bf y}-{\bf z}_2) } \frac{\big\langle  \hat{J}_a^\mu(x^+,{\bf z}_1) \hat{J}_b^\nu(y^+,{\bf z}_2)  \big\rangle_A}{{\bf q}_{1}^2 {\bf q}_{2}^2}.
	\end{align} 
	
	Using the result of \cref{eq:current_correlator}, we arrive at
	\begin{align}
		\label{eq:aux12}
		\Bigg\langle A_a^\mu\left(\vec{b} + \frac{\vec{r}}{2}\right) A_b^\nu\left(\vec{b} - \frac{\vec{r}}{2}\right) \Bigg\rangle_A
		=
		\delta^{ab} \delta(r^+)
		&\int_{{\bf P}, {\bf Q}}
		e^{i {\bf P} \cdot {\bf r}}
		\frac{\Gamma^\mu_{\sigma_1 \sigma_2}\left(P_q^-, {\bf P}; P_q^-,  -\frac{\textbf{Q}}{2} \right)
			\Gamma^\nu_{\sigma_2 \sigma_1}\left(P_q^-, \frac{\textbf{Q}}{2}; P_q^-, {\bf P}\right)}{2 \left({\bf P} + \frac{\bf Q}{2} \right)^2 \left({\bf P} - \frac{\bf Q}{2} \right)^2}
		\nonumber \\  \times &
			\int_{\textbf{B}} e^{i {\bf Q} \cdot ({\bf b} - {\bf B})} \mu^2(b^+, \textbf{B}).
	\end{align} 
	Here, we drop the hat notation for field operator from now on.
	
	Further simplification is possible under the assumption of a homogeneous nucleus, $\mu(\vec{x}) = \mu^2$. Introducing the transverse color charge density $\tilde{\mu}^2=\int_{x^+ }\mu^2(x^+) = L^+ \mu^2$, where $L^+$ represents the light-cone longitudinal width of the nucleus, \cref{eq:aux12} can be simplified to give
	\begin{align}\label{eq:field_correlation}
		&\big\langle A^\mu\left(\vec{x} \right) A^\nu\left(\vec{y}\right) \big\rangle_A
		=
		\delta^{ab} \delta(x^+ - y^+)
		\frac{\tilde{\mu}^2}{L^+}
		G^{\mu \nu} ({\bf x} - {\bf y}),
	\end{align} 
	where we define
	\begin{align}\label{eq:green_function}
		G^{\mu \nu} ({\bf r}) = \int_{{\bf P}}
		e^{i {\bf P} \cdot {\bf r}}
		\frac{\Gamma^\mu_{\sigma_1 \sigma_2}\left(P_q^-, {\bf P}; P_q^-,  0 \right)
			\Gamma^\nu_{\sigma_2 \sigma_1}\left(P_q^-, 0; P_q^-, {\bf P}\right)}{2 {\bf P}^4}.
	\end{align}
	
	\Cref{eq:field_correlation} is a key result of this work, as it allows us to calculate the correlation of the classical gluon background field generated by a large nucleus composed of quasi-independent quarks. In the following sections, we will leverage this equation to compute the nuclear DIS cross section at sub-eikonal accuracy.
	
	Prior to proceeding with our analysis, let us examine the scaling behavior of the correlator with respect to the nucleus rapidity $\omega$, $\gamma = e^{\omega}$. As established in the preceding section, the partonic longitudinal momentum scales with the nucleus rapidity as $P_q^- \propto \gamma$. Therefore, we can deduce the following scaling with $\gamma$ from \cref{eq:gamma}:
	\begin{align}
		G^{--} \propto 1, \qquad
		G^{-i} \propto \gamma^{-1}, \qquad
		G^{-+},  G^{ij} \propto \gamma^{-2}, \qquad
		G^{+i} \propto \gamma^{-3}, \qquad
		G^{++} \propto \gamma^{-4}.
	\end{align}
	
	As we will see in the next section, at sub-eikonal level, i.e., at $\mathcal{O}(\gamma^{-1})$, the only components of the field correlator that contribute are $--$, $-i$, $i-$ and $ij$. In this case, the non-trivial components of~\cref{eq:green_function} that we need are the following:
	\begin{align}
		\label{eq:Gmm}
		G^{--}({\bf r}) &= \int_{{\bf P}} e^{i {\bf P} \cdot {\bf r}}
		\frac{1}{{\bf P}^4},
		\\
		\label{eq:Gmi}
		G^{i-}({\bf r}) &= \frac{1}{2 P_q^-}\int_{{\bf P}} e^{i {\bf P} \cdot {\bf r}}
		\frac{{\bf P}^i}{{\bf P}^4},
		\\
		\label{eq:Gij}
		G^{ij}({\bf r}) &= \frac{1}{(2 P_q^-)^2}\int_{{\bf P}} e^{i {\bf P} \cdot {\bf r}}
		\frac{{\bf P}^i {\bf P}^j + \epsilon^{i m} \epsilon^{j n} {\bf P}^m {\bf P}^n}{{\bf P}^4}.
	\end{align}
	
	On the other hand, due to Lorentz contraction, $L^+ \propto \gamma^{-1}$ and $\delta(x^+-y^+) \propto \gamma$. This leads to the following scaling behavior for the color field correlators:
	\begin{align}
		\langle A_a^-(x^+,{\bf x}) A_b^-(y^+, {\bf y}) \rangle_A \propto \gamma^2, \qquad \langle A_a^-(x^+,{\bf x}) {\bf A}_b^j(y^+, {\bf y}) \rangle_A \propto \gamma, \qquad \langle {\bf A}_a^i(x^+,{\bf x}) {\bf A}_b^j(y^+, {\bf y}) \rangle_A \propto \gamma^0.
	\end{align}
	This observed scaling aligns with our expectations. Under a boost in the left direction, the gauge field transforms as $A^\mu = \Lambda^{\mu}_\nu A_0^\nu = ( \gamma A_0^-, \gamma^{-1} A_0^+, {\bf A}_0^i )$, with $A^\mu_0$ the background field in the rest frame of the nucleus.

\section{DIS dijet production beyond the eikonal approximation}
\label{sec:DIS_Neik}
	
	In this section, we review the calculation of dijet production in nuclear deep inelastic scattering (DIS) beyond the eikonal approximation within the CGC framework, presented in \cite{Altinoluk:2022jkk}. We employ the dipole picture of high-energy nuclear DIS, where the production of a forward quark-antiquark dijet arises from the splitting of a virtual photon $\gamma^*$ into a quark-antiquark dipole followed by its scattering with the nucleus. We work in a frame where the target's transverse momentum vanishes. The incoming electron has momentum $k_e^\mu$, colliding with a nucleus with momentum $P_A^\mu$. The interaction is mediated by a virtual photon with momentum $q^\mu = k_e'^\mu - k_e^\mu$, where $k_e'^\mu$ represents the final momentum of the electron. The virtual photon subsequently decays into a quark-antiquark pair with individual momenta $k_1^\mu$ and $k_2^\mu$, respectively. The quark and antiquark interact with the classical background field, $A^\mu(x)$, generated by the color sources within the nucleus.
	
	Within the single-photon exchange approximation, the DIS process is described by the product of a leptonic tensor, characterizing the virtual photon emission by the electron, and a hadronic tensor, characterizing the virtual photon's interaction with the target. After integrating over the azimuthal angle of the scattered electron, the hadronic tensor is projected onto the cross section of the $\gamma^* + A$ interaction:
	\begin{align}
		\label{eq:DIS_xsection}
		\frac{d \sigma^{e^- + A \to e + q  \bar{q} + X }}{d W^2 dQ^2 d^2 {\bf k}_1  d^2 {\bf k}_2 d \eta_1  d \eta_2}
		=
		\frac{\alpha_{\rm em} }{\pi Q^2 s y} \left[
		\left(1-y+\frac{y^2}{2}\right)
		\frac{d \sigma^{\gamma^*_T + A \to q  \bar{q} + X}}{ d^2 {\bf k}_1  d^2 {\bf k}_2 d \eta_1  d \eta_2}
		+
		(1-y)
		\frac{d \sigma^{\gamma^*_L + A \to q  \bar{q} + X}}{ d^2 {\bf k}_1  d^2 {\bf k}_2 d \eta_1  d \eta_2}
		\right].
	\end{align}
	Here, $y = \frac{W^2+Q^2-m_{\rm N}^2}{s-m_{\rm N}^2}$ is the inelasticity, $m_{\rm N} $ is the nucleon mass, $Q^2=-q^2$ is the photon virtuality, $\sqrt{s}$ is the center-of-mass (CoM) energy per nucleon of the $e^-+A$ system and $W$ is the CoM energy per nucleon of the $\gamma^*+A$ system. 
	
	The cross-section for $q\bar{q}$ production in a $\gamma^* + A$ collision, extending beyond the eikonal approximation, can be expressed as follows:
	\begin{align}
		\frac{d \sigma^{\gamma^*_\lambda + A \to q  \bar{q} + X}}{d^2 {\bf k}_1  d^2 {\bf k}_2 d \eta_1  d \eta_2}
		=
		\frac{d \sigma^{\gamma^*_\lambda + A \to q  \bar{q} + X}}{d^2 {\bf k}_1  d^2 {\bf k}_2 d \eta_1  d \eta_2} \Bigg|_{\rm Eik.}
		+
		\frac{d \sigma^{\gamma^*_\lambda + A \to q  \bar{q} + X}}{d^2 {\bf k}_1  d^2 {\bf k}_2 d \eta_1  d \eta_2} \Bigg|_{\rm NEik.}
		+ \mathcal{O}(\gamma^{-2}),
	\end{align}
	where $\lambda = L, T$ denotes the polarization of the virtual photon, either longitudinal ($L$) or transverse ($T$). The first term represents the eikonal approximation, which assumes an infinitely large rapidity difference between the virtual photon and the target. The second term introduces the $\gamma^{-1}$ order correction, denoted as NEik., which accounts for the first order deviations from the eikonal approximation.
	
	\begin{figure}[h!]
		\centering
		\includegraphics[scale=0.5]{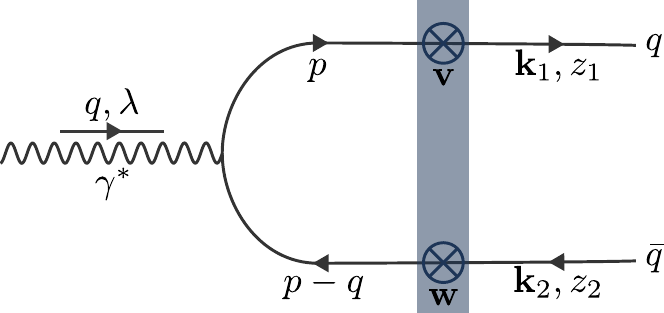}
		\caption{Effective eikonal amplitude diagram for $q \bar{q}$ production in nuclear DIS. A virtual photon with polarization $\lambda$ and momentum $q^\mu$ splits into a quark-antiquark pair. The color dipole interacts with the nuclear target via light-like Wilson lines. This eikonal interaction is represented by the blue rectangle.}
		\label{fig:dipole_eikonal}
	\end{figure}
	
	\Cref{fig:dipole_eikonal} shows the effective diagram for calculating the scattering amplitude for $q \bar{q}$ production in $\gamma^*+A$ collisions within the eikonal approximation. The incoming right-moving virtual photon undergoes splitting into a quark-antiquark dipole. This splitting process is analytically calculable using perturbative Quantum Electrodynamics (QED). On the other hand, the subsequent multiple interactions of the dipole with the background field generated by the target require non-perturbative treatment and are expressed in terms of light-like Wilson lines in the fundamental representation of SU($N_c$):
	\begin{align}
		\mathcal{U}_{[x^+,y^+]} ({\bf z}) = \mathcal{P}_+ 
		\exp\left\{  
		-ig \int_{y^+}^{x^+} dz^+ A^-(z^+, {\bf z})
		\right\},
	\label{eq:Wilson_line}
	\end{align}
	where $\mathcal{P}_+$ is the path order operator and, due to the infinite rapidity difference between the projectile and target, only the longitudinal component of the field contributes to the scattering. \Cref{eq:Wilson_line} represents the eikonal propagation of a fermion in a gluonic background field from longitudinal point $y^+$ to $x^+$. 
	
	Henceforth, by combining these elements, the cross section reads~\cite{Dominguez:2011wm}
	\begin{align}\label{eq:xsectionL_eik}
		\frac{d \sigma^{\gamma^*_\lambda + A \to q  \bar{q} + X}}{ d^2 {\bf k}_1  d^2 {\bf k}_2 d \eta_1  d \eta_2} \Bigg|_{\rm Eik.}
		&=
		\int_{{\bf v}, {\bf v}', {\bf w}, {\bf w}'} e^{i{\bf k}_1 \cdot ({\bf v}'-{\bf v})+i{\bf k}_2 \cdot ({\bf w}'-{\bf w})} 
		\mathcal{C}_\lambda({\bf w}'-{\bf v}', {\bf w}-{\bf v})
		\nonumber \\ & \hskip1cm \times
		\Big[
		Q({\bf w}', {\bf v}', {\bf v}, {\bf w}) - d({\bf w}', {\bf v}') - d({\bf v}, {\bf w}) + 1
		\Big],
	\end{align}
	where
	\begin{align}
		\mathcal{C}_L({\bf r}_1, {\bf r}_2) &= 
		\sum_f
		\frac{8 N_c \alpha_{\rm em} e_f^2 Q^2}{(2 \pi)^6} \delta_z  z_1^3 z_2^3 K_0(\epsilon_f |{\bf r}_1| )
		K_0(\epsilon_f |{\bf r}_2| ),
		\\
		\mathcal{C}_T({\bf r}_1, {\bf r}_2) &= 
		\sum_f
		\frac{2 N_c \alpha_{\rm em} e_f^2}{(2 \pi)^6} \delta_z  z_1 z_2 
		\left[
		m_f^2 + (z_1^2+z_2^2) \partial_{{\bf r}_1^j} \partial_{{\bf r}_2^j}
		\right]
		K_0(\epsilon_f |{\bf r}_1| )
		K_0(\epsilon_f |{\bf r}_2| ).
	\end{align}
	Here, we introduce the notation $\delta_z \equiv \delta(z_1 + z_2 - 1)$, where the variables $z_{1,2} = k_{1,2}^+/q^+$ represent the fraction of +-momentum of the photon carried by the (anti)quark. We also define $\epsilon_f^2 = m_f^2 + z_1 z_2 Q^2$, where $m_f$ denotes the mass of the quark with flavor $f$ and fractional charge $e_f$. 	\Cref{eq:xsectionL_eik} is factorized into two components: the factor $\mathcal{C}_\lambda$, which, aside from constant factors, represents the probability of splitting, and the dipole and quadrupole functions
	\begin{align}
		d({\bf v}, {\bf w}) &= \frac{1}{N_c} \Big\langle {\rm Tr}  \left[\mathcal{U}({\bf v}) \mathcal{U}^\dagger({\bf w})\right] \Big\rangle,
		\qquad
		\hskip0.4cm
		Q({\bf w}',{\bf v}',{\bf v},{\bf w}) = \frac{1}{N_c} \Big\langle {\rm Tr}  \left[\mathcal{U}({\bf w}') \mathcal{U}^\dagger({\bf v}')  \mathcal{U}({\bf v}) \mathcal{U}^\dagger({\bf w})\right] \Big\rangle,
		\label{eq:dipole}
	\end{align}
	which account for the eikonal interaction of the quark and antiquark with the nucleus. In this paper, we adopt the convention that when the particle propagates through the entire extent $L^+$ of the target (as is always occurs in the eikonal limit due to Lorentz contraction), the Wilson line is written as $\mathcal{U}({\bf x}) \equiv \mathcal{U}_{[L^+/2,-L^+/2]}({\bf x})$.
	
	At next-to-eikonal (NEik) accuracy, the classical background field loses its infinite contraction and acquires a finite light-cone width, denoted by $L^+$. Furthermore, the components $A^+$ and ${\bf A}^i$ along with the $x^-$ dependence of the field become non-zero. As previously mentioned, this manuscript excludes non-eikonal corrections stemming from both the $x^-$ dependence and the $A^+$ component of the field, as the latter only contributes at next-to-next-to-eikonal accuracy. In this case, the field correlators derived in~\cref{eq:field_correlation} within the covariant gauge ($\partial_\mu A^\mu = 0$) are equally valid in the light-cone gauge ($A^+ = 0$). This equivalence eliminates the need for gauge transformations when employing the correlators, as the results of~\cite{Altinoluk:2022jkk} were obtained in the light-cone gauge. Consequently, our study solely concentrates on next-to-eikonal corrections arising from the finite target width and the non-vanishing transverse components of the field.
	
%	\textcolor{red}{\bf P.A. : I would eliminate the following paragraph because the equation that we are using for the gauge transformation doesn't have an eikonal expansion: ($z^-$ scales as $\gamma$ while $A^+$ as $1/\gamma$.). Instead I would use the modification that I wrote above.}
%	
%	\textcolor{red}{
%	As previously mentioned, however, this manuscript will not consider non-eikonal corrections arising from the $x^-$ dependence. Moreover, the derived equations employed the light-cone gauge ($A^+ = 0$)\footnote{While the field correlators in~\cref{eq:field_correlation} were derived in the Lorentz gauge ($\partial_\mu A^\mu = 0$), their validity extends to the $A^+ = 0$ gauge at next-to-eikonal accuracy. This arises from the fact that the gauge transformation to the light-cone gauge becomes trivial at the eikonal order considered in this paper:
%	\begin{align}
%		\Omega (\vec{x}) = \mathcal{P} \exp \Bigg\{ 
%		i g \int_{x^-}^{\infty} dz^- A^+(z^-, {\bf x})
%		\Bigg\} = \mathbb{1} + \mathcal{O}({\gamma^{-2}}).
%	\end{align}}.}
	
	\begin{figure}[h!]
		\centering
		\includegraphics[scale=0.5]{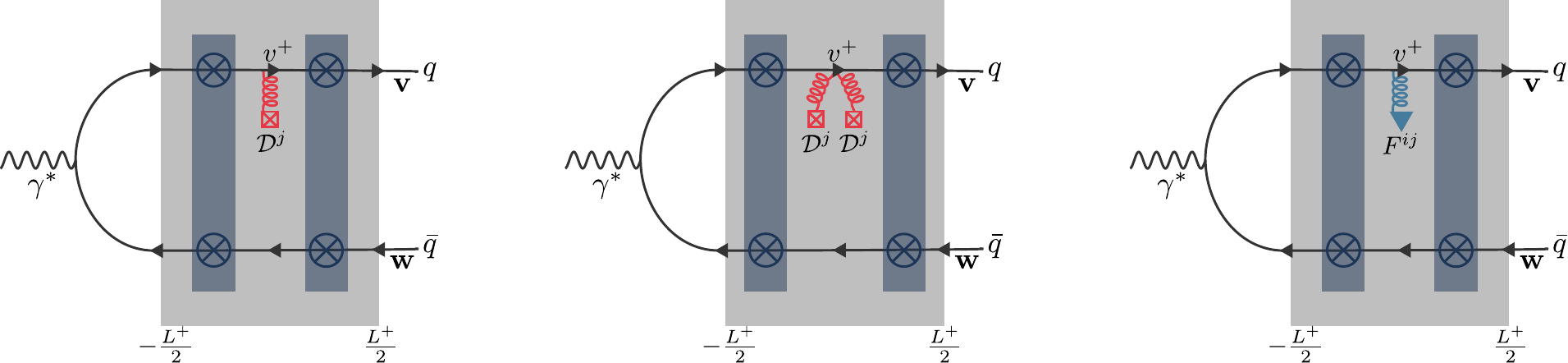}
		\caption{Effective NEik amplitude diagrams for $q \bar{q}$ production in nuclear deep inelastic scattering (DIS). The diagrams depict scenarios where the virtual photon splits into a quark-antiquark pair before entering the nucleus. After splitting, either the quark or antiquark (illustrated only for the quark here) propagates eikonally in the background field up to a longitudinal point $v^+$. At this point, insertions of the transverse covariant derivative or strength tensor occurs. Subsequently, the particle propagates eikonally again.}
		\label{fig:dipole_Neikonal}
	\end{figure}
	
\begin{figure}[h!]
		\centering
		\includegraphics[scale=0.5]{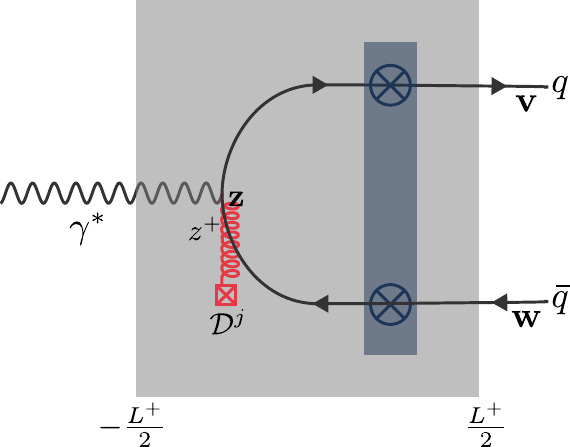}
		\caption{Effective NEik amplitude diagram for $q \bar{q}$ production in nuclear deep inelastic scattering (DIS). The diagram depicts the scenario where the virtual photon splits into a quark-antiquark pair inside the nucleus.}
		\label{fig:dipole_Neikonal_inside}
	\end{figure}

	Due to the medium's finite width, the photon can split either inside or before encountering it. In~\cref{fig:dipole_Neikonal}, we show a subset of effective diagrams contributing to the NEik amplitude when the photon splits before entering the medium. The blue rectangle represents the eikonal scattering of the partons (either quark or antiquark) with the medium. The non-eikonal corrections arise from inserting either the transverse covariant derivative $D_{{\bf z}^j}(v^+) = {\partial_{{\bf z}^j}} - i g {\bf A}^j(v^+,{\bf z})$, or the transverse field strength tensor $F_{ij}(z^+,{\bf z}) = \partial_{{\bf z}^i} {\bf A}_j(z^+,{\bf z}) - \partial_{{\bf z}^j} {\bf A}_i(z^+,{\bf z}) -ig[{\bf A}_i(z^+,{\bf z}),{\bf A}_j(z^+,{\bf z})]$, at a longitudinal point $v^+$ along the parton's path.
	On the other hand,~\cref{fig:dipole_Neikonal_inside} shows the case where the photon splits inside the target. In this case, the only type of next-to-eikonal corrections appears through a single transverse covariant derivative insertion in the longitudinal point where the photon splits.
	
	All in all, the non-eikonal corrections originating from these diagrams have been computed in \cite{Altinoluk:2022jkk}. The final result for the NEik corrections to the cross section for $q\bar{q}$ production in a $\gamma_\lambda^*+A$ collision is:
	\begin{align}\label{eq:xsection_NEik}
		&\frac{d \sigma^{\gamma^*_\lambda + A \to q  \bar{q} + X}}{ d^2 {\bf k}_1  d^2 {\bf k}_2 d \eta_1  d \eta_2} \Bigg|_{\rm NEik.} =
		\frac{1}{q^+} 
		{\rm Re}\int_{{\bf v}, {\bf w}, {\bf v}', {\bf w}'} e^{i{\bf k}_1 \cdot ({\bf v}'-{\bf v})+i{\bf k}_2 \cdot ({\bf w}'-{\bf w})} 
		\mathcal{C}_\lambda({\bf w}'-{\bf v}', {\bf w}-{\bf v})
		\nonumber \\ & \times
		\Bigg\{ 
		\frac{1}{z_1}
		\left[ \frac{{\bf k}_2^j-{\bf k}_1^j}{2} + \frac{i}{2} \partial_{{\bf w}^j}  \right] 
		\left[ Q_j^{(1)}({\bf w}',{\bf v}',{\bf v}_*,{\bf w}) - d_j^{(1)}({\bf v}_*,{\bf w})\right] 
		- \frac{i}{z_1} \left[ Q^{(2)}({\bf w}',{\bf v}',{\bf v}_*,{\bf w}) - d^{(2)}({\bf v}_*,{\bf w})  \right]
		\nonumber \\ & \hskip8pt -
		\frac{1}{z_2}
		\left[\frac{{\bf k}_2^j-{\bf k}_1^j}{2} - \frac{i}{2} \partial_{{\bf v}^j}  \right] 
		\left[ Q_j^{(1)}({\bf v}',{\bf w}',{\bf w}_*,{\bf v})^\dagger - d_j^{(1)}({\bf w}_*,{\bf v})^\dagger \right] 
		- \frac{i}{z_2} \left[ Q^{(2)}({\bf v}',{\bf w}',{\bf w}_*,{\bf v})^\dagger - d^{(2)}({\bf w}_*,{\bf v})^\dagger  \right]
		\Bigg\}
		\nonumber \\ & +
		\delta^{\lambda T}
		\frac{d \sigma^{\rm trans.}}{ d\Pi},
	\end{align}
	where we have introduced the factor $\delta^{\lambda T}$ which is 1 when $\lambda = T$ and 0 when $\lambda = L$, since the function $\frac{d \sigma^{\rm trans.}}{ d\Pi}$, to be defined below, only contributes to the cross section when the photon is transversely polarized. 
	
	\Cref{eq:xsection_NEik} gives the $\mathcal{O}(\gamma^{-1})$ correction to the eikonal cross section given in~\cref{eq:xsectionL_eik}. Indeed, we can see that the only parameter in~\cref{eq:xsection_NEik} that scales with $\gamma$ under a boost is $q^+$, leading to the expect $\gamma^{-1}$ scaling. Now, the non-eikonal interaction of the quark and antiquark with the background field is accounted for by the following modifications of the dipole and quadrupole functions:
	\begin{align}
		d_j^{(1)}({\bf v}_*,{\bf w}) &= \frac{1}{N_c} \Big\langle {\rm Tr}  \left[\mathcal{U}_j^{(1)}({\bf v}) \mathcal{U}^\dagger({\bf w})\right] \Big\rangle,
		\qquad
		Q_j^{(1)}({\bf w}',{\bf v}',{\bf v}_*,{\bf w}) = \frac{1}{N_c} \Big\langle {\rm Tr}  \left[\mathcal{U}({\bf w}') \mathcal{U}^\dagger({\bf v}')  \mathcal{U}_j^{(1)}({\bf v}) \mathcal{U}^\dagger({\bf w})\right] \Big\rangle,
		\label{eq:dipole1}
		\\
		d^{(2)}({\bf v}_*,{\bf w}) &= \frac{1}{N_c} \Big\langle {\rm Tr}  \left[\mathcal{U}^{(2)}({\bf v}) \mathcal{U}^\dagger({\bf w})\right] \Big\rangle,
		\qquad
		Q^{(2)}({\bf w}',{\bf v}',{\bf v}_*,{\bf w}) = \frac{1}{N_c} \Big\langle {\rm Tr}  \left[\mathcal{U}({\bf w}') \mathcal{U}^\dagger({\bf v}')  \mathcal{U}^{(2)}({\bf v}) \mathcal{U}^\dagger({\bf w})\right] \Big\rangle,
		\label{eq:dipole2}
	\end{align}
	where
	\begin{align}
		\label{eq:WL1}
		\mathcal{U}_j^{(1)}({\bf z}) &= \int_{-\frac{L^+}{2}}^{\frac{L^+}{2}} dv^+ \mathcal{U}_{[\frac{L^+}{2},v^+]}({\bf z}) \overleftrightarrow{D_{{\bf z}^j}}(v^+) \mathcal{U}_{[v^+,-\frac{L^+}{2}]}({\bf z}),
		\\
		\label{eq:WL2}
		\mathcal{U}^{(2)}({\bf z}) &= \int_{-\frac{L^+}{2}}^{\frac{L^+}{2}} dv^+ \mathcal{U}_{[\frac{L^+}{2},v^+]}({\bf z}) \overleftarrow{D_{{\bf z}^j}}(v^+)\overrightarrow{D_{{\bf z}^j}}(v^+) \mathcal{U}_{[v^+,-\frac{L^+}{2}]}({\bf z}).
	\end{align}
	In the context of next-to-eikonal corrections,~\cref{eq:WL1,eq:WL2} describe the propagation of a fermion through a medium with finite width, as depicted in the first and second diagrams of Fig.~\ref{fig:dipole_Neikonal}, respectively. Both equations capture the fermion's interactions with the background field but differ in the nature of the non-eikonal interactions at point $v^+$.
	
	In~\cref{eq:WL1}, the fermion undergoes eikonal propagation from the medium's beginning ($-L^+/2$) to $v^+$. At $v^+$, a non-eikonal interaction occurs through the insertion of $\overleftrightarrow{D_{{\bf z}^j}}(z^+) = {\partial_{{\bf z}^j}} - \overleftarrow{\partial_{{\bf z}^j}} - 2ig {\bf A}^j(z^+,{\bf z})$. Thus, the fermion  interacts with the transverse gradient of the longitudinal field and the transverse field itself. Following this, the fermion experiences another eikonal propagation until reaching the medium's end ($L^+/2$).~\Cref{eq:WL2} describes a similar scenario with a distinct non-eikonal interaction. Here, the interaction is represented by $\overleftarrow{D_{{\bf z}^j}}(v^+)\overrightarrow{D_{{\bf z}^j}}(v^+)$, where $\overleftarrow{D_{{\bf z}^j}}(z^+) = \overleftarrow{\partial_{{\bf z}^j}} + i g {\bf A}^j(z^+,{\bf z})$. In this case the fermion can interact with two gradients of the field as well as two transverse field insertions at the same longitudinal point.
	
	Moreover, the part of the cross section that only contributes for a transversely polarized photon is given by
	\begin{align}\label{eq:trans_contribution}
		\frac{d \sigma^{\rm trans.}}{ d\Pi} &=
		\frac{1}{q^+} 
		{\rm Re}\int_{{\bf v}, {\bf w}, {\bf v}', {\bf w}'} e^{i{\bf k}_1 \cdot ({\bf v}'-{\bf v})+i{\bf k}_2 \cdot ({\bf w}'-{\bf w})} 
		\mathcal{D}_{T,1}^{ij}({\bf w}'-{\bf v}', {\bf w}-{\bf v})
		\nonumber \\ & \hskip1cm \times
		\Bigg[\frac{i}{z_1} \left[ Q_{ij}^{(3)}({\bf w}',{\bf v}',{\bf v}_*,{\bf w}) - d_{ij}^{(3)}({\bf v}_*,{\bf w})  \right]
		+\frac{i}{z_2} \left[ Q_{ij}^{(3)}({\bf v}',{\bf w}',{\bf w}_*,{\bf v})^\dagger - d_{ij}^{(3)}({\bf w}_*,{\bf v})^\dagger  \right]\Bigg]
		\nonumber \\ &
		-
		\frac{1}{q^+} 
		{\rm Im}\int_{{\bf z}, {\bf v}', {\bf w}'} e^{i{\bf k}_1 \cdot ({\bf v}'-{\bf z})+i{\bf k}_2 \cdot ({\bf w}'-{\bf z})} 
		\mathcal{D}_{T,2}^{j}({\bf w}'-{\bf v}')
		\nonumber \\ & \hskip1cm \times
		\int_{-\frac{L^+}{2}}^{\frac{L^+}{2}} dz^+ \Bigg\langle  
		\frac{1}{N_c} {\rm Tr} \Big[\mathcal{U}({\bf w}') \mathcal{U}^\dagger({\bf v}') - 1 \Big]
		\Big[\mathcal{U}_{\left[\frac{L^+}{2}, z^+\right]}({\bf z})
		\overleftrightarrow{D_{{\bf z}^j}}(z^+) \mathcal{U}_{\left[\frac{L^+}{2}, z^+\right]}^\dagger({\bf z})
		- 
		\frac{1}{2}
		\mathcal{U}({\bf z})
		\overleftrightarrow{\partial_{{\bf z}^j}} \mathcal{U}^\dagger({\bf z})
		\Big]
		\Bigg\rangle
		,
	\end{align}
	where
	\begin{align}
		&\mathcal{D}_{T,1}^{ij}({\bf r}_1, {\bf r}_2) = 
		\sum_f
		\frac{2 N_c \alpha_{\rm em} e_f^2}{(2 \pi)^6} \delta_z  z_1 z_2 
		(z_1-z_2) \partial_{{\bf r}_1^i} \partial_{{\bf r}_2^j}
		K_0(\epsilon_f |{\bf r}_1|)
		K_0(\epsilon_f |{\bf r}_2|),
		\\ &
		\mathcal{D}_{T,2}^{j}({\bf r}) = 
		\sum_f
		\frac{N_c \alpha_{\rm em} e_f^2}{2 (2 \pi)^5} \delta_z [1+(z_2-z_1)^2] \partial_{{\bf r}^j}
		K_0(\epsilon_f |{\bf r}|).
	\end{align}  
	
	In first term of~\cref{eq:trans_contribution}, the quark and antiquark pair is generated before entering the medium and their interaction with the background field is accounted by the following modifications of the dipole and quadrupole operators:
	\begin{align}
		d_{ij}^{(3)}({\bf v}_*,{\bf w}) &= \frac{1}{N_c} \Big\langle {\rm Tr}  \left[\mathcal{U}_{ij}^{(3)}({\bf v}) \mathcal{U}^\dagger({\bf w})\right] \Big\rangle,
		\qquad
		Q_{ij}^{(3)}({\bf w}',{\bf v}',{\bf v}_*,{\bf w}) = \frac{1}{N_c} \Big\langle {\rm Tr}  \left[\mathcal{U}({\bf w}') \mathcal{U}^\dagger({\bf v}')  \mathcal{U}_{ij}^{(3)}({\bf v}) \mathcal{U}^\dagger({\bf w})\right] \Big\rangle,
		\label{eq:dipole3}
	\end{align}
	where 
	\begin{align}
		\mathcal{U}_{ij}^{(3)}({\bf z}) &= \int_{-\frac{L^+}{2}}^{\frac{L^+}{2}} dv^+ \mathcal{U}_{[\frac{L^+}{2},v^+]}({\bf z}) g F_{ij}(v^+,{\bf z}) \mathcal{U}_{[v^+,-\frac{L^+}{2}]}({\bf z}).
		\label{eq:WL3}
	\end{align}
	\Cref{eq:WL3} defines a modified Wilson line, capturing the case where the non-eikonal corrections arise from the insertion of the transverse strength tensor, as shown in the third diagram of~\cref{fig:dipole_Neikonal}.	The second term of~\cref{eq:trans_contribution} represents the scenario where the photon splits inside the medium, as depicted in~\cref{fig:dipole_Neikonal_inside}. 
	
\section{DIS dijet production in the dilute limit beyond the eikonal approximation}
\label{sec:DIS_dilute}
	
	This section aims to calculate the cross section for dijet production in nuclear DIS beyond the eikonal approximation using the CGC framework given in~\cref{eq:xsectionL_eik,eq:xsection_NEik}. Specifically, we will employ the dilute limit approximation, which offers significant computational advantages while remaining applicable under certain regions of the kinematic space. In particular, when the transverse momenta of the $q \bar{q}$ pair exceeds the saturation scale of the target nucleus, $Q_s$, the interaction regime switches from dense to dilute. In this scenario, the gluon density within the nucleus becomes sufficiently low that individual gluon scatterings with the $q \bar{q}$ pair dominate, rather than collective interactions with the entire Color Glass Condensate. 
	
	The main advantage for utilizing the dilute limit approximation is that the physics become perturbative and we can approximate the Wilson line as 
	\begin{align}\label{eq:WL_expansion}
		\mathcal{U}_{[x^+,y^+]} ({\bf z}) =  
		1 
		-ig \int_{y^+}^{x^+} dz^+ A^-(z^+, {\bf z})
		- g^2 \int_{y^+}^{x^+} dz_1^+ \int_{y^+}^{z_1^+} dz_2^+  A^-(z_1^+, {\bf z}) A^-(z_2^+, {\bf z}) + \mathcal{O}(g^3),
	\end{align}
	so that the interaction of the partons with the medium can be approximated by the exchange of two gluons instead of a multi-gluon interaction. This allows us to directly use the non-eikonal field correlator derived in~\cref{sec:field_correlator}. In contrast, exploring the dense limit would necessitate: (i) develop a method to resum the non-eikonal correlators, accounting for the multi-gluon exchange contributions that become essential in the dense regime, and (ii) take into account the non-linear terms of the classical YM equation. 
	
	Therefore, due to the substantial increase in complexity associated with the dense limit beyond the eikonal approximation, in this paper we study the cross section in the kinematic regime where ${\bf k}_1^2, {\bf k}_2^2 \gtrsim Q_s^2$. In this case the dilute limit approximation is valid and presents a more computationally efficient and tractable approach while maintaining sufficient accuracy for our purposes.
	
\subsection{DIS dijet production at eikonal accuracy in the dilute limit}
	
	Our calculation begins with determining the cross section at eikonal accuracy. To achieve this, we require a model for the dipole and quadrupole operators defined in \cref{eq:dipole}. Leveraging the expansion of the Wilson line from \cref{eq:WL_expansion}, we express the dipole function at second order in the coupling constant as follows:
	\begin{align}
		\label{eq:aux13}
		d({\bf x},{\bf y}) &=
		1 - \frac{g^2}{N_c}
		\int_{-\frac{L^+}{2}}^{\frac{L^+}{2}} z^+ \int_{-\frac{L^+}{2}}^{z^+} dz_1^+
		\nonumber \\ & \hskip2cm \times
		{\rm Tr}\Big[
		\langle A^-(z^+, {\bf x}) A^-(z_1^+, {\bf x}) \rangle
		+ \langle A^-(z_1^+, {\bf y}) A^-(z^+, {\bf y}) \rangle
		- 2 \langle A^-(z^+, {\bf x}) A^-(z_1^+, {\bf y}) \rangle
		\Big].
	\end{align}
	
	Next, we employ the expression for the field correlator provided in~\cref{eq:field_correlation} to write
	\begin{align}\label{eq:aux7}
		\frac{g^2}{N_c} \big\langle {\rm Tr} \left[A^\mu\left(\vec{x} \right) A^\nu\left(\vec{y}\right)\right] \big\rangle
		= 2 \pi Q_s^2 \frac{\delta(x^+-y^+)}{L^+}
		G^{\mu \nu} ({\bf x} - {\bf y}).
	\end{align}
	Here, we have defined the saturation momentum as
	\begin{align}
		\label{eq:saturation_momentum}
		 Q_s^2 \equiv \frac{g^2}{2 \pi} C_F \tilde{\mu}^2 \approx \frac{g^4 A C_F}{4 \pi^2 R_A^2},
	\end{align}
	where in the last step we used the fact the for a homogeneous circular nucleus, the transverse color charge density squared is $\tilde{\mu}^2 \approx g^2 A /(2 \pi R_A^2)$, where $R_A \sim A^{1/3}$ fm is the radius of the nucleus.
	
	By using the above expression, Eq.~\eqref{eq:aux13} can be written as follows\footnote{We note that by resumming higher order correlators, we arrive at the well known expression within the MV model:
		\begin{align}
			d({\bf x}, {\bf y}) = \exp\left\{ - \frac{Q_s^2}{4} ({\bf x} - {\bf y})^2 \ln \frac{1}{|{\bf x} - {\bf y}|\Lambda_{\rm QCD}} \right\},
		\end{align}
		when $|{\bf x} - {\bf y}| \lesssim Q_s$.
	}:
	\begin{align}\label{eq:dip_aux}
		d({\bf x},{\bf y}) =
		1- 2 \pi Q_s^2
		\left[
		G^{--} (0)
		- G^{--} ({\bf x} - {\bf y})
		\right].
	\end{align}
	Similarly, in the dilute limit of the CGC, the quadrupole function is expressed as
	\begin{align}\label{eq:quad_aux}
		Q({\bf w}', {\bf v}', {\bf v}, {\bf w}) 
		= 1 - 2 \pi Q_s^2
		\Big[&
		2 G^{--} (0)
		- G^{--} ({\bf v}' - {\bf v})
		- G^{--} ({\bf v} - {\bf w})
		+ G^{--} ({\bf v}' - {\bf w})
		\nonumber \\ &
		+ G^{--} ({\bf w}' - {\bf v})	
		- G^{--} ({\bf w}' - {\bf v}')
		- G^{--} ({\bf w}' - {\bf w})
		\Big].
	\end{align}
	
\begin{figure}[h!]
		\centering
		\includegraphics[scale=0.5]{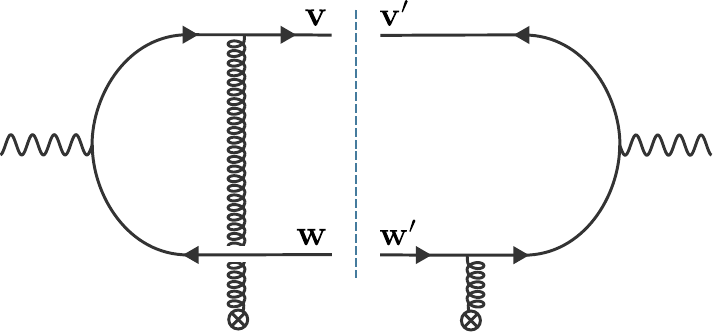}
		\caption{One of the four diagrams contributing to the eikonal differential cross section for quark-antiquark production in DIS in the dilute limit. The circle with a cross represents a longitudinal source.}
		\label{fig:dilute_diagrams_eikonal}
	\end{figure}

	All in all, the eikonal term of the cross section, presented in~\cref{eq:xsectionL_eik}, reads
	\begin{align}\label{eq:aux9}
		\frac{d \sigma^{\gamma^*_\lambda + A \to q  \bar{q} + X}}{d^2 {\bf k}_1 d \eta_1 d^2 {\bf k}_2 d \eta_2} \Bigg|_{\rm Eik.}
		&=
		2 \pi Q_s^2
		\int_{{\bf v}, {\bf v}', {\bf w}, {\bf w}'} e^{i{\bf k}_1 \cdot ({\bf v}'-{\bf v})+i{\bf k}_2 \cdot ({\bf w}'-{\bf w})} 
		\mathcal{C}_\lambda({\bf w}'-{\bf v}', {\bf w}-{\bf v})
		\nonumber \\ & \times
		\Big[
		G^{--} ({\bf v}' - {\bf v})
		- G^{--} ({\bf v}' - {\bf w})
		- G^{--} ({\bf w}' - {\bf v})	
		+ G^{--} ({\bf w}' - {\bf w})
		\Big].
	\end{align}
	It is clear that this equation solely accounts for the interaction of the quark (or antiquark) with a single gluon in both the amplitude and complex amplitude, as depicted in \cref{fig:dilute_diagrams_eikonal}. 
	
	Unlike in the dense limit,~\cref{eq:aux9} can be solved exactly and it does not depend on the non-perturbative scale $\Lambda_{\rm QCD}$. This can be explicitly seen by performing the following change of variables:
	\begin{align}
		\label{eq:COV}
		{\bf v} = {\bf B} - \frac{{\bf r} + {\bf R}}{2}, \qquad
		{\bf v}' = {\bf B} - \frac{{\bf b} - {\bf R}}{2}, \qquad
		{\bf w} = {\bf B} + \frac{{\bf r} - {\bf R}}{2}, \qquad
		{\bf w}' = {\bf B} + \frac{{\bf b} + {\bf R}}{2}, \qquad
	\end{align}
	which simplifies~\cref{eq:aux9} to
	\begin{align}\label{eq:aux8}
		\frac{d \sigma^{\gamma^*_\lambda + A \to q  \bar{q} + X}}{d^2 {\bf k}_1 d \eta_1 d^2 {\bf k}_2 d \eta_2} \Bigg|_{\rm Eik.}
		=
		2 \pi Q_s^2 S_\perp
		\int_{{\bf b}, {\bf r}} \left(
		e^{i {\bf b} \cdot {\bf k}_2} - e^{-i {\bf b} \cdot {\bf k}_1}
		\right)
		\left(
		e^{-i {\bf r} \cdot {\bf k}_2} - e^{i {\bf r} \cdot {\bf k}_1}
		\right) 
		\mathcal{C}_\lambda({\bf r}, {\bf b})
		\int_{\bf R} e^{i {\bf R} \cdot ({\bf k}_1+{\bf k}_2)}
		G^{--} ({\bf R}).
	\end{align}
	Here, the factor $S_\perp \equiv \int d^2 {\bf B}$ arises, since we are assuming a homogeneous and therefore translationally invariant target.~\Cref{eq:aux8} can be solved analytically using the definition of $G^{--}$ in~\cref{eq:Gmm} as well as the following identities:
	\begin{align}\label{eq:bessel_int}
		\int_{\bf x} e^{i {\bf x} \cdot {\bf k}} K_0(\epsilon_f |{\bf x}|) = 2 \pi \frac{1}{{\bf k}^2 + \epsilon_f^2}, \qquad
		\int_{\bf x} e^{i {\bf x} \cdot {\bf k}} \partial_{{\bf x}^j} K_0(\epsilon_f |{\bf x}|) = -2 \pi i \frac{ {\bf k}^j}{{\bf k}^2 + \epsilon_f^2}.
	\end{align}
	
	By solving the transverse integrals using~\cref{eq:bessel_int}, we obtain the expression for the differential cross-section for quark and antiquark production at eikonal accuracy:
	\begin{align}
		\frac{d \sigma^{\gamma^*_L + A \to q  \bar{q} + X}}{d^2 {\bf k}_1 d \eta_1 d^2 {\bf k}_2 d \eta_2} \Bigg|_{\rm Eik.}
		&=
		\sum_f C^{(f)}_L
		\frac{Q^2 ({\bf k}_1^2-{\bf k}_2^2)^2}{({\bf k}_1+{\bf k}_2)^4 ({\bf k}_1^2 + \epsilon_f^2)^2 ({\bf k}_1^2 + \epsilon_f^2)^2},
		\label{eq:xSectionL_Eik}
		\\
		\frac{d \sigma^{\gamma^*_T + A \to q  \bar{q} + X}}{d^2 {\bf k}_1 d \eta_1 d^2 {\bf k}_2 d \eta_2} \Bigg|_{\rm Eik.}
		&=
		\sum_f C^{(f)}_T
		\frac{1}{({\bf k}_1+{\bf k}_2)^4}
		\left[
		m_f^2 \frac{({\bf k}_1^2-{\bf k}_2^2)^2}{({\bf k}_2^2 + \epsilon_f^2)^2 ({\bf k}_1^2 + \epsilon_f^2)^2} + (z_1^2+z_2^2) 
		\left( \frac{{\bf k}_2^i}{{\bf k}_2^2 + \epsilon_f^2} + \frac{{\bf k}_1^i}{{\bf k}_1^2 + \epsilon_f^2} \right)^2
		\right],
		\label{eq:xSectionT_Eik}
	\end{align}
	where we have defined the following dimensionless parameters:
	\begin{align}
		C^{(f)}_L = \frac{8 N_c \alpha_{\rm em} e_f^2 Q_s^2 S_\perp}{(2 \pi)^3} \delta_z  z_1^3 z_2^3,
		\qquad
		C^{(f)}_T =
		\frac{2 N_c \alpha_{\rm em} e_f^2 Q_s^2 S_\perp}{(2 \pi)^3} \delta_z z_1 z_2.
	\end{align}
	
	It follows, then, that in the dilute regime of CGC, the eikonal differential cross-section for dijet production in DIS is analytical. Moreover, unlike in the dense limit within the MV model, the infrared regulator $\Lambda_{\rm QCD}$ is not required. While the results presented in~\cref{eq:xSectionL_Eik,eq:xSectionT_Eik} thus do include non-perturbative modes, this does not significantly affect the results provided that ${\bf k}_1^2, {\bf k}_2^2 \gtrsim Q_s^2 \gg \Lambda_{\rm QCD}^2$.
	
\subsection{DIS dijet production beyond eikonal accuracy in the dilute limit}
	
	We now turn to calculating the next-to-eikonal corrections to the differential cross-section using the same approach as before. As outlined in~\cref{sec:DIS_Neik}, non-eikonal corrections can be categorized into four distinct types based on how the parton interacts with the medium. If the splitting occurs before the photon enters the medium, these corrections include: (i) a single covariant derivative insertion, (ii) a double covariant derivative insertion, and (iii) a transverse strength field insertion. We also need to consider the contribution when (iv) the splitting occurs inside the medium. In the following, we compute all these contributions in the dilute limit of the CGC.

\subsubsection{Single covariant derivative contribution}

	The case in which the parton interacts with the background field with a single transverse covariant derivative insertion is accounted for by the modification of the Wilson line given in~\cref{eq:WL1}. In the dilute limit, it can be written as follows:
	\begin{align}
		\mathcal{U}_j^{(1)}({\bf v})
		&=
		\int_{-\frac{L^+}{2}}^{\frac{L^+}{2}} dv^+ \Bigg[
		-ig \int_{-\frac{L^+}{2}}^{v^+} dz_1^+ \left(  \partial_{{\bf v}^j} A^-(z_1^+,{\bf v}) - \partial_{{\bf v}^j} A^-(v^+,{\bf v}) \right)
		- 2ig {\bf A}^j(v^+,{\bf v})
		\nonumber \\ & +
		g^2 \int_{-\frac{L^+}{2}}^{v^+} dz_1^+ \int_{-\frac{L^+}{2}}^{z_1^+} dz_2^+ \partial_{{\bf v}^j} \left(
		A^-(v^+,{\bf v}) A^-(z_1^+,{\bf v}) - A^-(z_1^+,{\bf v}) A^-(z_2^+,{\bf v})
		\right)
		\nonumber \\ & -
		2g^2 \int_{-\frac{L^+}{2}}^{v^+} dz_1^+ \Big({\bf A}^j(v^+,{\bf v}) A^-(z_1^+,{\bf v}) 
		+
		 A^-(v^+,{\bf v}) {\bf A}^j(z_1^+,{\bf v})
		\Big)
		\Bigg] + \mathcal{O}(g^3).
	\end{align}
	
	Using this expansion, it is straightforward to compute in the dilute limit the modifications of the dipole and quadrupole operators given in~\cref{eq:dipole1}:
	\begin{align}
		d_j^{(1)} ({\bf v}_*, {\bf w}) &= 4 \pi Q_s^2 \left[ G^{-j}({\bf w} - {\bf v}) - G^{j-}(0) \right],
		\\
		Q_j^{(1)}({\bf w}',{\bf v}',{\bf v}_*,{\bf w}) &= 4 \pi Q_s^2 \left[ 
		G^{-j}({\bf w} - {\bf v}) + G^{-j}({\bf v}' - {\bf v}) -G^{-j}({\bf w}' - {\bf v})- G^{j-}(0) \right].
	\end{align}
	
\begin{figure}[h!]
		\centering
		\includegraphics[scale=0.5]{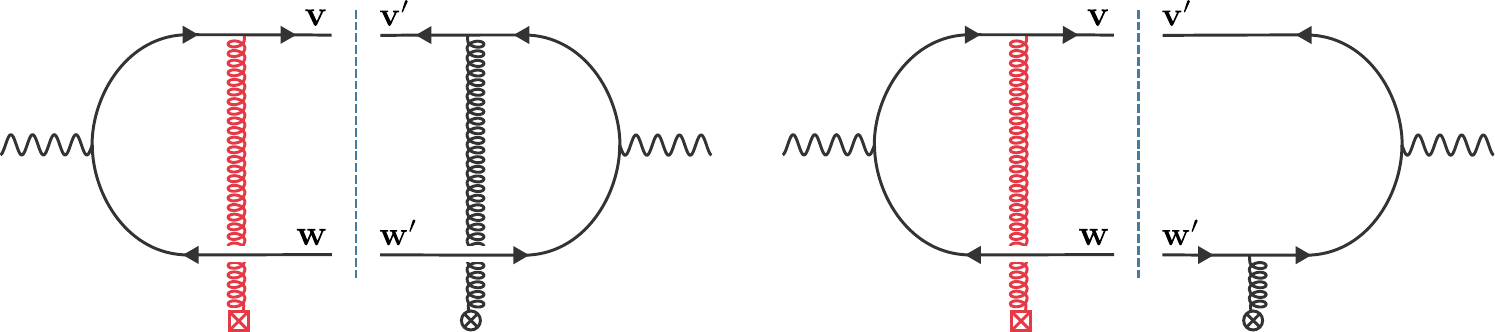}
		\caption{Two of the four diagrams contributing to the next-to-eikonal differential cross section for quark-antiquark production in DIS with a single insertion. The red curly line represents the exchange of a transverse gluon, while the black line represents a longitudinal gluon. The red square represents a transverse source of color charge, and the circle with a cross represents a longitudinal source. The diagrams with the antiquark interacting with the transverse gluon are not shown but are analogous to the ones presented here.
		}
		\label{fig:dilute_diagrams_NEik}
	\end{figure}

	In conclusion, we note that the next-to-eikonal corrections in~\cref{eq:xsection_NEik} solely arise from the difference between the modified quadrupole and dipole functions:
	\begin{align}
		\label{eq:aux15}
		Q_j^{(1)}({\bf w}',{\bf v}',{\bf v}_*,{\bf w}) - d_j^{(1)} ({\bf v}_*, {\bf w}) = 4 \pi Q_s^2 \left[ 
		G^{-j}({\bf v}' - {\bf v}) -G^{-j}({\bf w}' - {\bf v}) \right].
	\end{align}
	This type of non-eikonal correction, independent of the gradient of the longitudinal field component, describes the scenario where a parton interacts with the transverse field in the amplitude, while the partons in the complex amplitude interact with the longitudinal component.~\Cref{fig:dilute_diagrams_NEik} illustrates this, where the red curly line represents the transverse gluon and the black line represents the longitudinal gluon. We only show the case where the quark interacts with the transverse component, but there are two additional diagrams where the antiquark interacts with the transverse field.
	
\subsubsection{Double covariant derivative contribution}

	The case where the parton interacts with the background field through a double transverse covariant derivative insertion is accounted for by the modification of the Wilson line given in~\cref{eq:WL2}. In the dilute limit, this modified Wilson line can be written as follows:
	\begin{align}
		\mathcal{U}^{(2)}({\bf v}) &= - g^2 \int_{-\frac{L^+}{2}}^{\frac{L^+}{2}} dv^+ \Bigg[
		\int_{-\frac{L^+}{2}}^{v^+} dz_1^+ \int_{-\frac{L^+}{2}}^{z_1^+} dz_2^+ 
		\partial_{{\bf v}^j} A^-(z_1^+,{\bf v}) \partial_{{\bf v}^j} A^-(z_2^+,{\bf v})
		\nonumber \\ &
		+
		\int_{-\frac{L^+}{2}}^{v^+} dz_1^+ \Big(\partial_{{\bf v}^j} A^-(v^+,{\bf v}) {\bf A}^j(z_1^+,{\bf v})
		- {\bf A}^j(v^+,{\bf v}) \partial_{{\bf v}^j} A^-(z_1^+,{\bf v})
		\Big)
		- {\bf A}^j(v^+,{\bf v}) {\bf A}^j(v^+,{\bf v})
		\Bigg]+\mathcal{O}(g^3).
	\end{align}
	
	Since $\mathcal{U}^{(2)} \propto g^2$, the modified dipole and quadrupole functions defined in~\cref{eq:dipole2} are given by
	\begin{align}
		Q^{(2)}({\bf w}',{\bf v}',{\bf v}_*,{\bf w}) = d^{(2)}({\bf v}_*,{\bf w}) + \mathcal{O} (g^3) = \langle \mathcal{U}^{(2)} ({\bf v}) \rangle + \mathcal{O} (g^3)  .
	\end{align}
	This implies that these operators do not contribute to the NEik differential cross-section in the dilute limit:
	\begin{align}
		Q^{(2)}({\bf w}',{\bf v}',{\bf v}_*,{\bf w}) - d^{(2)}({\bf v}_*,{\bf w}) = 0.
	\end{align}
	
\subsubsection{Transverse strength tensor contribution}

	The case where the parton interacts with the background field through a transverse strength tensor insertion is accounted for by the modification of the Wilson line given in~\cref{eq:WL3}. In the dilute limit, this modified Wilson line can be written as follows:
	\begin{align}
		\mathcal{U}_{ij}^{(3)}({\bf v}) &=  \int_{-\frac{L^+}{2}}^{\frac{L^+}{2}} dv^+ \Bigg[
		g \Big(
		\partial_{{\bf v}^j} {\bf A}^i(v^+,{\bf v}) - \partial_{{\bf v}^i} {\bf A}^j(v^+,{\bf v}) - ig [{\bf A}^i(v^+,{\bf v}), {\bf A}^j(v^+,{\bf v})]
		\Big)
		 \\  & \hskip-1cm
		-ig^2 \int_{-\frac{L^+}{2}}^{v^+} dz_1^+ \Big(
		A^-(v^+,{\bf v}) [\partial_{{\bf v}^j} {\bf A}^i(z_1^+,{\bf v}) - \partial_{{\bf v}^i} {\bf A}^j(z_1^+,{\bf v})]
		+
		[\partial_{{\bf v}^j} {\bf A}^i(v^+,{\bf v}) - \partial_{{\bf v}^i} {\bf A}^j(v^+,{\bf v})] A^-(z_1^+,{\bf v})
		\Big)
		\Bigg]+\mathcal{O}(g^3).\nonumber
	\end{align}
	
	Up to order $g^2$, all terms contributing to the modification of the dipole and quadrupole functions (Eq.~\eqref{eq:dipole3}) involve correlators like 
	\begin{align}
		\label{eq:aux14}
		\langle  [\partial_{{\bf v}^j} {\bf A}^i(v^+,{\bf v}) - \partial_{{\bf v}^i} {\bf A}^j(v^+,{\bf v})] A^-(w^+,{\bf w}) \rangle  
		&\propto \partial_{{\bf v}^j} G^{i-}({\bf v} - {\bf w}) - \partial_{{\bf v}^i} G^{j-}({\bf v} - {\bf w}) 
		\nonumber \\ &
		= i \int_{\bf P} e^{i {\bf P} \cdot ({\bf x} - {\bf y})} \frac{{\bf P}^i {\bf P}^j - {\bf P}^j {\bf P}^i}{{\bf P}^4} = 0,
	\end{align}
	where in the last step we have used the definition of $G^{i-}$ given in~\cref{eq:Gmi},
	and
	\begin{align}
		\langle  [{\bf A}^i(v^+,{\bf v}), {\bf A}^j(v^+,{\bf v})] \rangle \propto G^{ij}(0) - G^{ji}(0)  = 0,
	\end{align}
	where we have used the fact that $G^{ij}$, given by~\cref{eq:Gij}, is symmetric in the transverse indices.
	
	Therefore, in the dilute limit and for a homogeneous target, corrections stemming from the field strength tensor insertion do not contribute to the cross section because
	\begin{align}
		Q_{ij}^{(3)}({\bf w}',{\bf v}',{\bf v}_*,{\bf w}) = d_{ij}^{(3)}({\bf v}_*,{\bf w}) = 0.
	\end{align}

	We note that the vanishing result in~\cref{eq:aux14} relies on the assumption of a homogeneous nucleus. For an inhomogeneous target with a non-trivial density distribution, these corrections are expected to become non-negligible. In fact, the magnitude of their contribution likely increases with the degree of inhomogeneity in the density distribution. Further investigation of such inhomogeneous scenarios, using more complex models for $\mu^2(b^+,{\bf b})$, could be crucial for achieving a more accurate understanding of these corrections and their impact on the overall calculations.

\subsubsection{Splitting inside contribution}
	
	Finally, we now analyze the corrections stemming from the case in which the photon splits in the $q{\bar q}$ dipole in a coordinate $z^+, {\bf z}$ inside the target. In this case we need to evaluate the following correlator:
	\begin{align}
		Q_{\rm in}({\bf w}', {\bf v}', {\bf z}) \equiv \int_{-\frac{L^+}{2}}^{\frac{L^+}{2}} dz^+ \Bigg\langle  
		\frac{1}{N_c} \Big[\mathcal{U}({\bf w}') \mathcal{U}^\dagger({\bf v}') - 1 \Big]
		\Big[\mathcal{U}_{\left[\frac{L^+}{2}, z^+\right]}({\bf z})
		\overleftrightarrow{D_{{\bf z}^j}}(z^+) \mathcal{U}_{\left[\frac{L^+}{2}, z^+\right]}^\dagger({\bf z})
		- 
		\frac{1}{2}
		\mathcal{U}({\bf z})
		\overleftrightarrow{\partial_{{\bf z}^j}} \mathcal{U}^\dagger({\bf z})
		\Big]
		\Bigg\rangle
	\end{align}
	
	Expanding the Wilson lines, it is possible to see that the contributions coming from the longitudinal field gradient vanish. Thus, after performing the longitudinal integrals this operator can be written as
	\begin{align}
		\label{eq:aux16}
		Q_{\rm in}({\bf w}', {\bf v}', {\bf z}) = 4 \pi Q_s^2 \left[ 
		G^{-j}({\bf v}' - {\bf z}) -G^{-j}({\bf w}' - {\bf z}) \right].
	\end{align}
	We see that this is the same result obtained in \cref{eq:aux15} but with ${\bf v} = {\bf w} = {\bf z}$. 
	
	\begin{figure}[h!]
		\centering
		\includegraphics[scale=0.5]{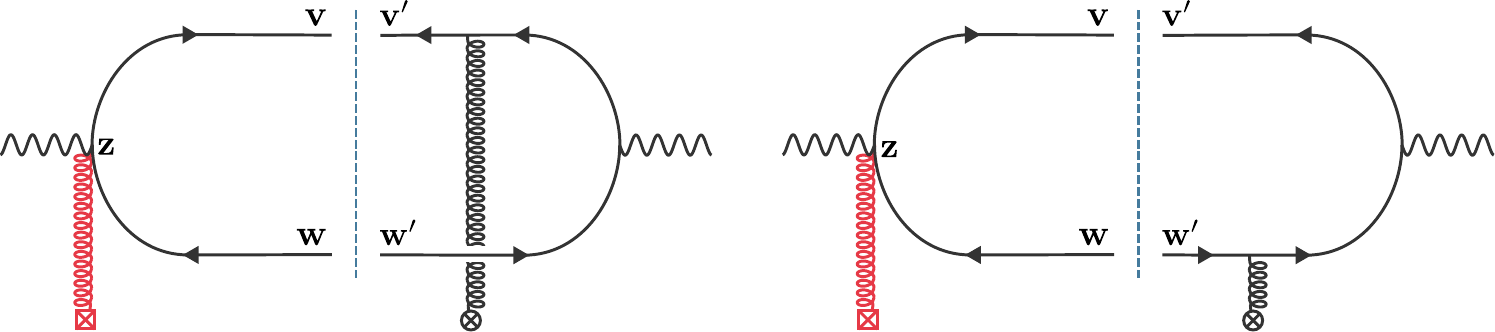}
		\caption{Diagrams contributing to the next-to-eikonal differential cross section for quark-antiquark production in DIS when the photon splits inside the target. The red curly line represents the exchange of a transverse gluon, while the black line represents a longitudinal gluon. The red square represents a transverse source of color charge, and the  circle with a cross represents a longitudinal source.
		}
		\label{fig:dilute_diagrams_NEik_inside}
	\end{figure}
	
\subsubsection{Summing up the contributions}	
	
	In summary, for a homogeneous target within the dilute limit, the next-to-eikonal corrections to the differential cross section are solely determined by the single covariant derivative contribution (\cref{eq:aux15}) and the contribution from in-medium splitting (\cref{eq:aux16}). These contributions are depicted in the diagrams presented in~\cref{fig:dilute_diagrams_NEik,fig:dilute_diagrams_NEik_inside}. 
	
	By substituting~\cref{eq:aux15,eq:aux16} into the NEik differential cross section definition (\cref{eq:xsection_NEik,eq:trans_contribution}) and utilizing the change of variables in~\cref{eq:COV}, we arrive at the following expression for the cross section:
	\begin{align}
		&\frac{d \sigma^{\gamma^*_\lambda + A \to q  \bar{q} + X}}{d^2 {\bf k}_1 d \eta_1 d^2 {\bf k}_2 d \eta_2} \Bigg|_{\rm NEik.}
		=
		\frac{2 \pi Q_s^2 S_\perp}{q^+}
		\int_{{\bf b}, {\bf r}, {\bf R}} 
		e^{i {\bf R} \cdot ({\bf k}_1+{\bf k}_2)}
		\mathcal{C}_\lambda({\bf r}, {\bf b})
		\nonumber \\ & \times
		\Bigg[
		e^{-i {\bf r} \cdot {\bf k}_2}
		\left(
		e^{i {\bf b} \cdot {\bf k}_2} - e^{-i {\bf b} \cdot {\bf k}_1}
		\right)
		\frac{({\bf k}_2 - {\bf k}_1)^j}{z_1}
		G^{-j} ({\bf R})
		+
		e^{-i {\bf r} \cdot {\bf k}_1}
		\left(
		e^{i {\bf b} \cdot {\bf k}_1} - e^{-i {\bf b} \cdot {\bf k}_2}
		\right)
		\frac{({\bf k}_1 - {\bf k}_2)^j}{z_2}
		G^{-j} ({\bf R})^*
		\Bigg]
		\nonumber \\ & \hskip3cm - 
		\delta^{\lambda T}
		 \frac{4 \pi Q_s^2 S_\perp}{q^+} {\rm Im}\int_{{\bf b}, {\bf R}} 
		e^{i {\bf R} \cdot ({\bf k}_1+{\bf k}_2)} \mathcal{D}_{T,2}^j ({\bf b}) 
		\left(
		e^{i {\bf b} \cdot {\bf k}_2} - e^{-i {\bf b} \cdot {\bf k}_1}
		\right)
		G^{-j}({\bf R}).
	\end{align}
	
	Finally, solving this integral is straightforward, by using the definition of $G^{- i}$ given in~\cref{eq:Gmi} together with the Bessel integral given in~\cref{eq:bessel_int}. Analogous to~\cref{eq:xSectionL_Eik,eq:xSectionT_Eik}, the result is also analytical and given by
	\begin{align}
		\frac{d \sigma^{\gamma^*_L + A \to q  \bar{q} + X}}{d^2 {\bf k}_1 d \eta_1 d^2 {\bf k}_2 d \eta_2} \Bigg|_{\rm NEik.}
		&=
		\frac{C_T}{2 q^+ P_q^-}
		\frac{Q^2 ({\bf k}_1^2-{\bf k}_2^2)^2}{({\bf k}_1+{\bf k}_2)^4 ({\bf k}_1^2 + \epsilon_f^2)^2 ({\bf k}_1^2 + \epsilon_f^2)^2}
		\left[\frac{{\bf k}_1^2 + \epsilon_f^2}{z_1}
		-\frac{{\bf k}_2^2 + \epsilon_f^2}{z_2}
		\right],
		\label{eq:xSectionL_NEik}
		\\
		\frac{d \sigma^{\gamma^*_T + A \to q  \bar{q} + X}}{d^2 {\bf k}_1 d \eta_1 d^2 {\bf k}_2 d \eta_2}
		\Bigg|_{\rm NEik.}
		&=
		\frac{C_T}{2 q^+ P_q^-}
		\frac{{\bf k}_1^2 - {\bf k}_2^2}{({\bf k}_1+{\bf k}_2)^4}
		\Bigg\{ 
		m_f^2 \frac{{\bf k}_1^2-{\bf k}_2^2}{({\bf k}_1^2 + \epsilon_f^2)^2 ({\bf k}_1^2 + \epsilon_f^2)^2} \Bigg[
		\frac{{\bf k}_1^2 + \epsilon_f^2}{z_1} - \frac{{\bf k}_2^2 + \epsilon_f^2}{z_2}
		\Bigg]
		\nonumber \\ & \hskip0cm +
		(z_1^2 + z_2^2) 
		\left( \frac{{\bf k}_2^i}{{\bf k}_2^2 + \epsilon_f^2} + \frac{{\bf k}_1^i}{{\bf k}_1^2 + \epsilon_f^2} \right)
		\Bigg[
		\frac{1}{z_1} \frac{{\bf k}_2^i}{{\bf k}_2^2 + \epsilon_f^2} +
		\frac{1}{z_2} \frac{{\bf k}_1^i}{{\bf k}_1^2 + \epsilon_f^2} +
		\frac{1}{z_1 z_2} \frac{{\bf k}_1^i + {\bf k}_2^i}{{\bf k}_1^2 - {\bf k}_2^2}
		\Bigg]
		\Bigg\}.
		\label{eq:xSectionT_NEik}
	\end{align}
	These equations represent the next-to-eikonal, $\mathcal{O}(\gamma^{-1})$, corrections to the differential cross section for dijet production in $\gamma^* + A$ collisions in the dilute limit of the nucleus. Indeed, these corrections are suppressed by $q^+ P_q^- \propto e^{\Delta Y}$, where $\Delta Y$ is the rapidity difference between the quarks inside the nucleus and the photon. Moreover, it is important to note that since~\cref{eq:xSectionL_NEik,eq:xSectionT_NEik} were computed in the dilute approximation, they are only valid in the regime where ${\bf k}_1^2, {\bf k}_2^2 \gtrsim Q_s^2$.
	
\subsection{The differential cross section at subeikonal order}
	
	To finalize the section, we sum the differential cross section at eikonal accuracy given in~\cref{eq:xSectionL_Eik,eq:xSectionT_Eik} with their NEik corrections given in~\cref{eq:xSectionL_NEik,eq:xSectionT_NEik}. This sum represents the subeikonal differential cross section, which incorporates the NEik corrections to the eikonal result. 
	Moreover, we express the suppression in the NEik contribution in terms of the COM energy per nucleon of the $\gamma^*+A$ system. In order to do so, we note that for a energetic nucleus we can write $W^2=(q+P_{\rm N})^2 \approx 2 N_c q^+ P_q^-$, where we have used the fact that all the quarks in the nucleon have the same longitudinal momentum, $P_{\rm N}^- = N_c P_q^-$. Therefore, the subeikonal differential cross section for a longitudinally and transversely polarized photon can be written as follows:
	\begin{align}\label{eq:xsectionL_subeik}
		\frac{d \sigma^{\gamma^*_L + A \to q  \bar{q} + X}}{d^2 {\bf k}_1 d \eta_1 d^2 {\bf k}_2 d \eta_2}
		&=
		\frac{d \sigma^{\gamma^*_L + A \to q  \bar{q} + X}}{d^2 {\bf k}_1 d \eta_1 d^2 {\bf k}_2 d \eta_2} \Bigg|_{\rm Eik.}
		\Bigg[  
		1 +
		\frac{N_c}{ W^2}
		\left(\frac{{\bf k}_1^2 + \epsilon_f^2}{z_1}
		- \frac{{\bf k}_2^2 + \epsilon_f^2}{z_2}\right)
		\Bigg]
		+ \mathcal{O}(\gamma^{-2}),
	\\
		\frac{d \sigma^{\gamma^*_T + A \to q  \bar{q} + X}}{d^2 {\bf k}_1 d \eta_1 d^2 {\bf k}_2 d \eta_2}
		&=
		\sum_f \frac{C_T^{(f)}}{({\bf k}_1+{\bf k}_2)^4}
		\Bigg\{
		m_f^2 \frac{({\bf k}_1^2-{\bf k}_2^2)^2}{({\bf k}_2^2 + \epsilon_f^2)^2 ({\bf k}_1^2 + \epsilon_f^2)^2} 
		\Bigg[  
		1 +
		\frac{N_c}{ W^2}
		\left(\frac{{\bf k}_1^2 + \epsilon_f^2}{z_1}
		- \frac{{\bf k}_2^2 + \epsilon_f^2}{z_2}\right)
		\Bigg]
		\nonumber \\ & \hskip-2cm +
		(z_1^2+z_2^2) 
		\left( \frac{{\bf k}_2^i}{{\bf k}_2^2 + \epsilon_f^2} + \frac{{\bf k}_1^i}{{\bf k}_1^2 + \epsilon_f^2} \right)
		\Bigg(
		\frac{{\bf k}_2^i}{{\bf k}_2^2 + \epsilon_f^2} \Bigg[1 + \frac{N_c}{ W^2} \frac{{\bf k}_1^2-{\bf k}_2^2}{z_1} \Bigg]
		+
		\frac{{\bf k}_1^i}{{\bf k}_1^2 + \epsilon_f^2} \Bigg[1 + \frac{N_c}{ W^2} \frac{{\bf k}_1^2-{\bf k}_2^2}{z_2} \Bigg]
		\nonumber \\ & \hskip4cm +
		\frac{N_c}{W^2} \frac{{\bf k}_1^i + {\bf k}_2^i}{z_1 z_2}
		\Bigg)
		\Bigg\}
		+ \mathcal{O}(\gamma^{-2}).
	\label{eq:xsectionT_subeik}
	\end{align}
	 These equations are the main result of this section and describe the differential cross section for $q {\bar q}$ production in DIS at subeikonal accuracy when the final transverse momentum of the partons are relatively large, ${\bf k}_1^2, {\bf k}_2^2 \gtrsim Q_s^2$. In the next section, we will analyze these solutions numerically to gain further insights into the dynamics of DIS in the subeikonal regime.
	
\section{Numerical results}
\label{sec:numerical_results}
		
	In this section, we perform a numerical analysis of the subeikonal differential cross sections,~\cref{eq:xsectionL_subeik,eq:xsectionT_subeik}, derived in the previous section, and we contrast them with the eikonal approximation. Within this analysis, the target nucleus is chosen as Au, with $A=197$ nucleons and characterized by a nuclear radius $R_A = 6.37$ fm which translates into a transverse area of $S_\perp = \pi R_A^2$. A saturation momentum $Q_s = 0.6$ GeV is taken. Given the present study's focus on qualitative insights on the effect of non-eikonal corrections, the small-$x$ evolution of the dipole and quadrupole operators is not considered.
	
	We focus on massless, $m_f = 0$, light quarks. This choice implies $\epsilon_f^2 = z_1 z_2 Q^2$, thereby eliminating any quark flavor dependence in the cross section except for the factor $\sum_f e_f^2 = \left(\frac{2}{3}\right)^2 + 2 \left(\frac{1}{3}\right)^2$. Additionally, the CoM energy per nucleon of the collision is fixed to be $\sqrt{s} = 90$ GeV, which corresponds to the top EIC energy for collisions on nuclear targets. We utilize the standard values for the number of colors ($N_c = 3$) and the fine structure constant ($\alpha_{\rm em} = 1/137$).
	
	The pseudorapidity ($\eta_{1,2}$) and longitudinal momentum fraction ($z_{1,2}$) of the dijet partons are linked by 
	$z_{1,2} = \frac{2 E_{\rm N}}{W^2} |\textbf{k}_{1,2}| e^{\eta_{1,2}},$ where $E_{\rm N}$ is the beam energy per nucleon. Given that $z_2 = 1 - z_1$, we investigate the asymmetry between $z_1$ and $z_2$ defining
	\begin{equation}
		\xi = \ln \frac{1-z_1}{z_1} = \ln \frac{|\textbf{k}_2|}{|\textbf{k}_1|} + \Delta \eta,
	\end{equation}
	where $\Delta \eta = \eta_2 - \eta_1$ is the pseudorapidity difference. This variable takes the value $\xi = 0$ for symmetric longitudinal momentum fractions ($z_1 = z_2 = 0.5$), and approaches $\pm \infty$ when $z_{1,2} = 0$.
	
	Furthermore, due to rotational symmetry, the cross section depends only on the azimuthal angle between the quark and antiquark defined as $\cos \phi_{q \bar{q}} = {\bf k}_1 \cdot {\bf k}_2 / k_{1 \perp} k_{2 \perp}$, with $k_{1,2 \perp} \equiv |{\bf k}_{1,2}|$. Hence, integrating over $z_2$ using the Dirac delta function, the phase space measure adopts the form $d \Pi = 2 \pi k_{1 \perp} d k_{1 \perp} k_{2 \perp} d k_{2 \perp} d \phi_{q \bar{q}} d \xi$.	
	
	\begin{figure}[h!]
		\centering
		\includegraphics[scale=0.7]{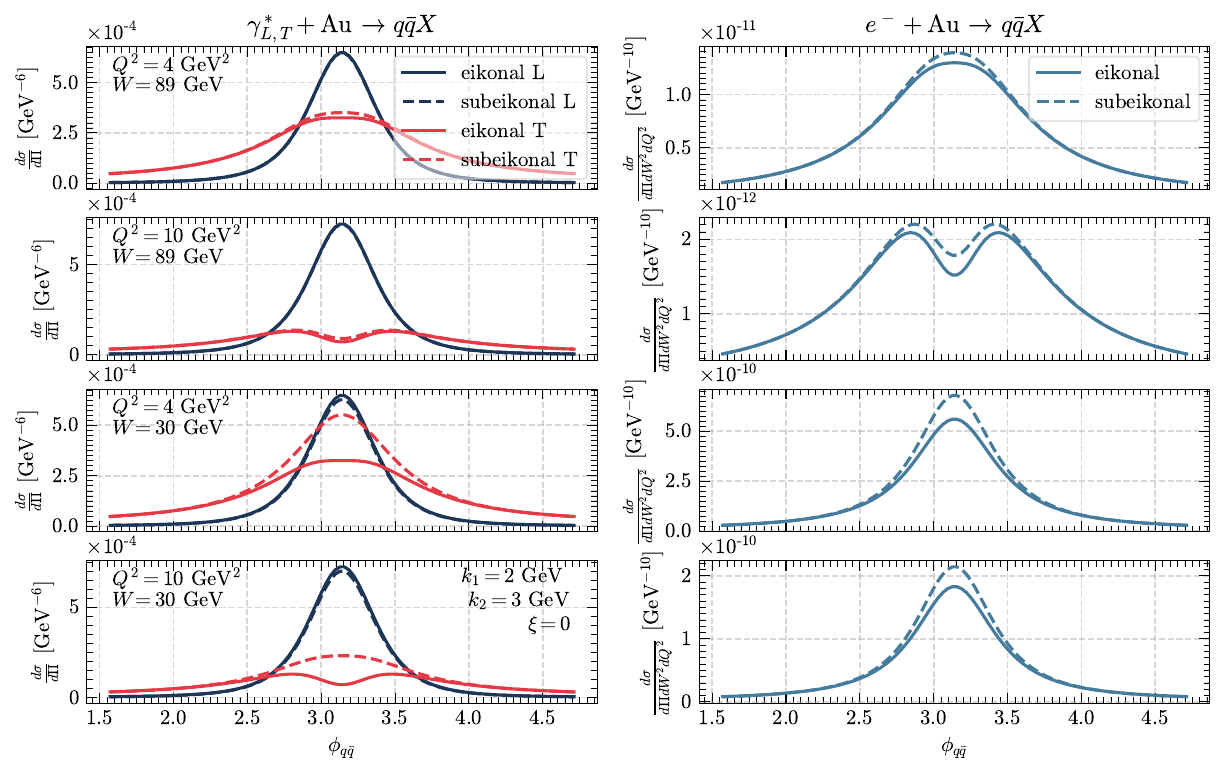}
		\caption{Differential cross section as a function of the angle between the quark and antiquark, $\phi_{q\bar{q}}$, at fixed values of $k_{1\perp} = 2$ GeV, $k_{2\perp} = 3$ GeV, and $\xi = 0$. {Left:} Transverse (red) and longitudinal (black) differential cross sections for the $\gamma^* + \text{Au}$ system. {Right:} Differential cross section per unit of $W^2$ and $Q^2$ for $e^- + \text{Au}$ collisions. Results are shown for different values of $Q^2$ and $W$, with the solid lines representing the eikonal result and the dashed lines representing the eikonal+subeikonal one.}
		\label{fig:xsection_phiqq}
	\end{figure}
	
	\Cref{fig:xsection_phiqq} showcases the impact of subeikonal corrections on the dijet angular distribution. We quantitatively compare the eikonal and (eikonal+)subeikonal differential cross sections as a function of the dijet angle ($\phi_{q\bar{q}}$) at fixed values of transverse momenta ($k_{1\perp} = 2$ GeV, $k_{2\perp} = 3$ GeV) and longitudinal momentum asymmetry ($\xi = 0$). This analysis is conducted for various combinations of center-of-mass energy ($W^2$) and momentum transfer ($Q^2$). Additionally, we quantify the subeikonal corrections in the cross section for the $e^- + A$ system defined in~\cref{eq:DIS_xsection}.
	
	Our observations reveal that the back-to-back peak centered around $\phi_{q\bar{q}} = \pi$ exhibits a broader profile in the very inelastic scattering regime ($W \approx \sqrt{s}$) compared to a less inelastic case ($W \ll \sqrt{s}$). Furthermore, for larger $Q^2$ values (corresponding to a higher inelasticity $y$), a double-peak structure emerges. This shows that the dominance of the back-to-back peak diminishes as momentum transfer increases. Notably, the non-eikonal corrections become more pronounced for smaller $W$ values, substantially modifying the back-to-back peak of the DIS cross section. Given the magnitude of these modifications, we anticipate that subeikonal corrections will hold sizable importance in future experimental measurements at the Electron-Ion Collider.
	
\subsection{Azimuthal anisotropy}
	
	Motivated by its relevance in the correlation limit of the Color Glass Condensate and to facilitate comparison with existing literature, we describe quark-antiquark momenta using the momentum imbalance, $\bold{\Delta} = \bold{k}_1 + \bold{k}_2$, and the relative momentum, $\bold{P} = z_2 \bold{k}_1 - z_1 \bold{k}_2$. $\bold{P}$ relates to the dijet pair's invariant mass as $M^2_{q \bar{q}} = P_\perp^2/(z_1 z_2)$, where $P_\perp = |{\bf P}|$, $\Delta_\perp = |{\bf \Delta}|$, and $\phi$ is the azimuthal angle difference between $\bold{P}$ and $\bold{\Delta}$. In these variables, the phase space measure simplifies to $d \Pi = 2 \pi P_\perp d P_\perp \Delta_\perp d \Delta_\perp d \phi d\xi$.
	
	For the $\gamma^* + A$ system\footnote{For the $e^-+A$ system, the harmonics expansion is analogous, with the key difference that the angle-averaged cross section is given in units of $Q^2$ and $W^2$. We denote this by $\tilde{D}_0(P_\perp, \Delta_\perp)$.}, the differential cross section can be expressed as a harmonics expansion
	\begin{align}
		\frac{d\sigma_\lambda^{\gamma^* A \to q \bar{q}X}}{d \Pi} &= D_{0, \lambda} (P_\perp, \Delta_\perp) \left[ 1 + 2 \sum_{n=1}^{\infty} v_{n, \lambda}(P_\perp, \Delta_\perp) \cos \phi \right],
	\end{align}
	where
	\begin{align}
		D_{n, \lambda} (P_\perp, \Delta_\perp) &= \int \frac{d \phi}{2 \pi} e^{in \phi} \frac{d\sigma^{\gamma_\lambda^* A \to q \bar{q}X}}{d \Pi}, \qquad
		v_{n, \lambda} = \frac{D_{n, \lambda}}{D_{0, \lambda}}.
	\end{align}
	Leveraging the expressions derived in~\cref{sec:DIS_dilute}, we analyze the angle-averaged differential cross section, $D_{0,\lambda}$, and the azimuthal harmonics $v_{n,\lambda}$ as functions of $P_\perp$, $\Delta_\perp$ and $\xi$.
	
	\begin{figure}[h!]
		\centering
		\includegraphics[scale=0.7]{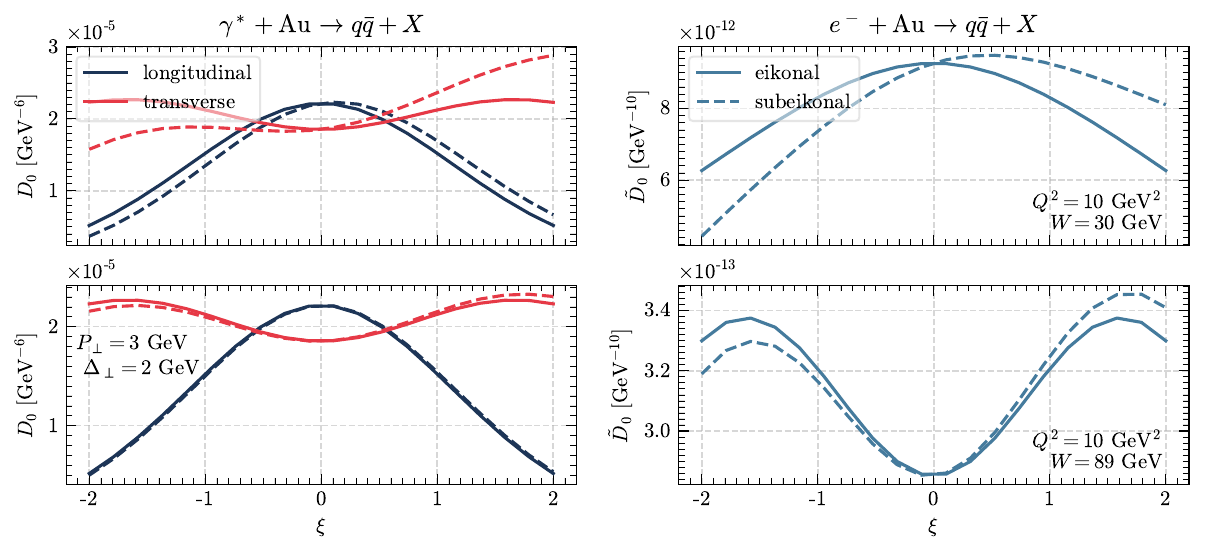}
		\caption{Angle-averaged dijet cross section as a function of the fraction of longitudinal momentum asymmetry, $\xi$, at fixed value of $P_\perp = 3$ GeV and $\Delta_\perp = 2$ GeV. Left: Transverse (red) and longitudinal (black) differential cross sections for the $\gamma^* + \text{Au}$ system; Right: Cross section for  $e^- + {\rm Au}$ collisions. Results are shown for different values of $Q^2$ and $W$,  with the solid line representing the eikonal result and the dashed lines including the non-eikonal corrections.}
		\label{fig:d0_vs_xi}
	\end{figure}

	Figure~\ref{fig:d0_vs_xi} shows the dependence of the angle-averaged cross section $D_{0,\lambda}$ with respect to $\xi$ at both eikonal and subeikonal accuracy. Fixed values of $P_\perp = 3$ GeV, $\Delta_\perp = 2$ GeV, and $Q^2 = 10$ GeV are employed. The first row displays results for relatively low inelasticity ($y = 0.11$), while the second row shows the case for high inelasticity ($y = 0.98$). We present the results for both $\gamma^* + \text{Au}$ and $e^- + \text{Au}$ collisions in separate panels.
	
	We observe that the eikonal cross section is symmetric with respect to $\xi$, while the subeikonal corrections break this symmetry. As expected, non-eikonal corrections are larger for smaller $W$ values but vanish when $\xi = 0$. This behavior can be analyzed analytically through~\cref{eq:xSectionL_NEik}, where we see that the non-eikonal terms, when $z_1 = z_2$, are proportional to ${\bf k}_1^2 - {\bf k}_2^2 \propto \cos \phi$ and therefore only modify odd harmonics.
	
\begin{figure}[h!]
		\centering
		\includegraphics[scale=0.7]{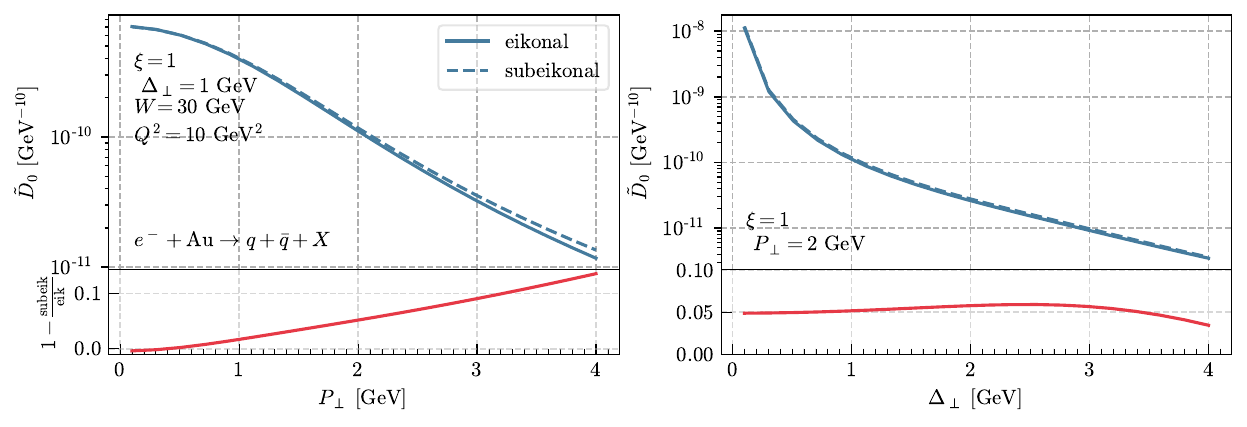}
		\caption{Left: Angle averaged dijet cross section as a function of $P_\perp$ at a fixed value of $\Delta_\perp = 1$ GeV and $\xi = 1$. Right: Angle averaged dijet cross section as a function of $\Delta_\perp$ at a fixed value of $P_\perp = 2$ GeV and $\xi = 1$. The solid line represents the eikonal result while the dashed line includes the non-eikonal corrections.}
		\label{fig:d0_vs_pperp}
	\end{figure}

	\Cref{fig:d0_vs_pperp} shows the dependence of the angle-averaged reduced cross section per unit of $W^2$ and $Q^2$, $\tilde{D}_{0,\lambda}$, for $e^- +{\rm Au}$ collisions, on $P_\perp$ and $\Delta_\perp$, at eikonal and subeikonal accuracy. The non-eikonal corrections exhibit a strong dependence on $P_\perp$, but only a weak variation with respect to $\Delta_\perp$. For $P_\perp \gtrsim 2.5$ GeV, $\Delta_\perp =1$ GeV, and $\xi = 1$, the non-eikonal corrections reach approximately 10 \%, while for $P_\perp = 1$ GeV and $\xi=1$, they amount to $\sim 5\%$, irrespective of the $\Delta_\perp$ value.
	
\begin{figure}[h!]
		\centering
		\includegraphics[scale=0.7]{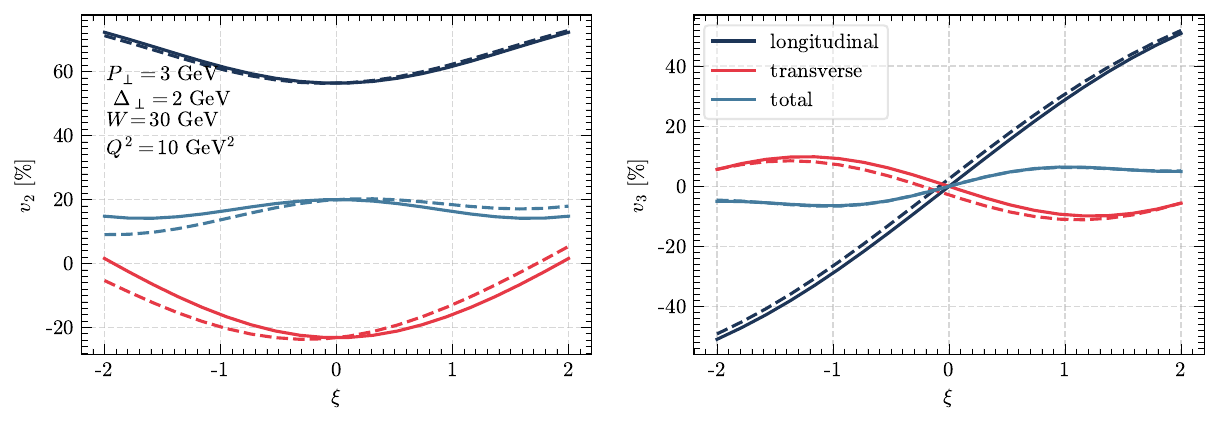}
		\caption{Left: Dependence of the elliptic harmonic, $v_2$, with respect to the longitudinal momenta asymmetry, $\xi$. Right: Dependence of the triangular harmonic, $v_3$, with respect to $\xi$. Both figures are for a fixed $P_\perp = 3$ GeV and $\Delta_\perp = 2$ GeV. We denote the harmonics for the $e^-+{\rm Au}$ collision as "total".}
		\label{fig:d2_vs_xi}
	\end{figure}

	\Cref{fig:d2_vs_xi} depicts the dependence of the elliptic ($v_2$) and triangular ($v_3$) harmonics on the longitudinal momentum asymmetry, $\xi$, at fixed $P_\perp = 3$ GeV and $\Delta_\perp = 2$ GeV. Similar to~\cref{fig:d0_vs_xi}, the non-eikonal corrections break the (anti)symmetry of the even (odd) azimuthal harmonics with respect to $\xi$. Notably, the subeikonal cross section exhibits non-zero odd harmonics even when the longitudinal momentum fractions are symmetric ($\xi = 0$), unlike the eikonal case. Importantly, in contrast to what is found in the correlation limit (see~\cref{sec:correlation_limit}), this approach reveals a dependence of both odd and even harmonics on the longitudinal momentum asymmetry.
	
\begin{figure}[h!]
		\centering
		\includegraphics[scale=0.7]{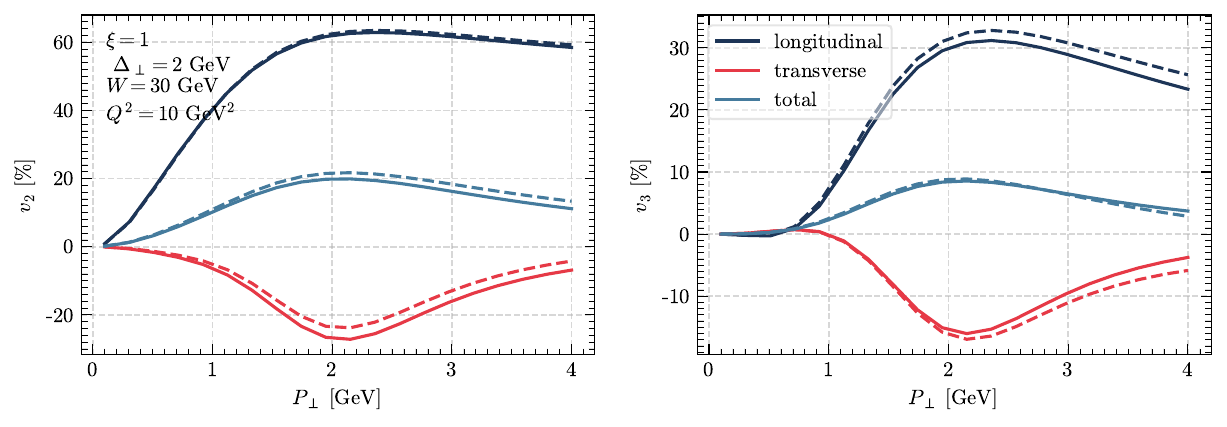}
		\caption{Left: Dependence of the elliptic harmonic, $v_2$, with respect to $P_\perp$. Right: Dependence of the triangular harmonic, $v_3$, with respect to $P_\perp$. Both figures are for fixed $\Delta_\perp = 2$ GeV and $\xi = 1$. We denote the harmonics for the $e^-+{\rm Au}$ collision as "total".}
		\label{fig:d2_vs_pperp}
	\end{figure}

	\Cref{fig:d2_vs_pperp} illustrates the dependence of the elliptic ($v_2$) and triangular ($v_3$) harmonics on the transverse momentum transfer, $P_\perp$, at fixed $\Delta_\perp = 2$ GeV and $\xi = 1$. Consistent with~\cref{fig:d0_vs_pperp}, significant corrections arise from non-eikonal effects at high $P_\perp$ values. As expected, both $v_2$ and $v_3$ exhibit a saturation behavior at large momenta.
	
\begin{figure}[h!]
		\centering
		\includegraphics[scale=0.7]{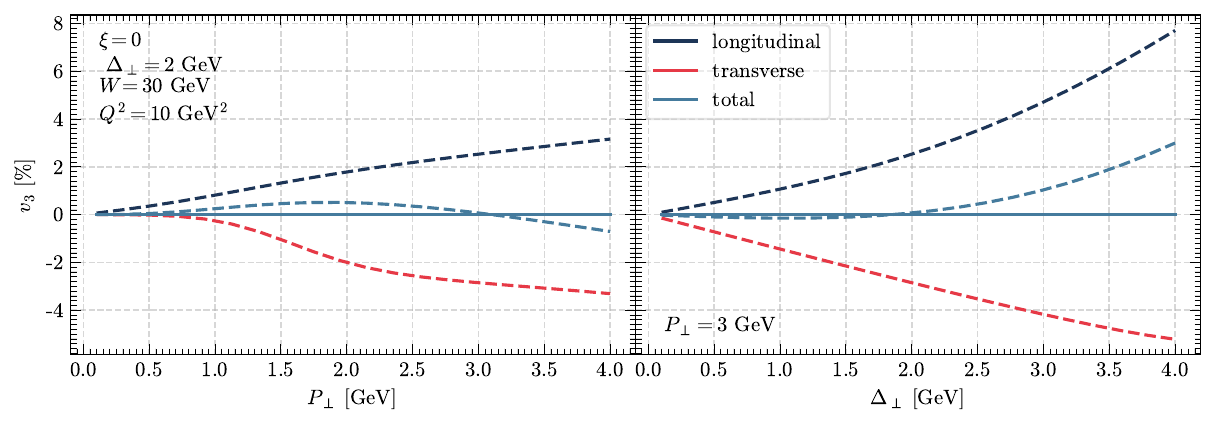}
		\caption{Left: Dependence of the triangular harmonic, $v_3$, with respect to $P_\perp$ at $\Delta_\perp = 2$ GeV. Right: Dependence of the triangular harmonic, $v_3$, with respect to $\Delta_\perp$ at $P_\perp = 3$ GeV. Both plots are for the case in which the quark-antiquark longitudinal momentum are the same, $\xi = 0$. We denote the harmonics for the $e^-+{\rm Au}$ collision as "total".}
		\label{fig:d3_vs_pperp}
	\end{figure}

	Furthermore,~\cref{fig:d3_vs_pperp} depicts the odd azimuthal harmonic as a function of $P_\perp$ and $\Delta_\perp$ for symmetric longitudinal momentum fractions ($\xi = 0$). Unlike the eikonal case, non-eikonal corrections generate non-zero odd harmonics, reaching approximately 5\%. Measuring these odd harmonics in nuclear DIS experiments therefore provides a direct probe of the non-eikonal contributions.
	
\subsection{Comparison with the dense limit and the correlation limit}
	
	To finalize, this section presents a comprehensive comparison of the eikonal angle-averaged cross section. We analyze the expression derived within the dilute limit of the CGC framework (\cref{eq:xSectionL_Eik,eq:xSectionT_Eik}) from~\cref{sec:DIS_dilute}. These expressions are compared to the dense limit introduced in~\cref{sec:dense_limit} and, owing to its analytical simplicity, to the correlation limit ($\Delta_\perp \ll P_\perp$) described in~\cref{sec:correlation_limit}.
	
	This comparative analysis serves two primary objectives: (i) it offers a stringent validation of the expressions derived within this manuscript, and (ii) it establishes quantitatively the kinematic regime of $P_\perp$ and $\Delta_\perp$ wherein the methodology employed in this study remains applicable.
	
\begin{figure}[h!]
		\centering
		\includegraphics[scale=0.6]{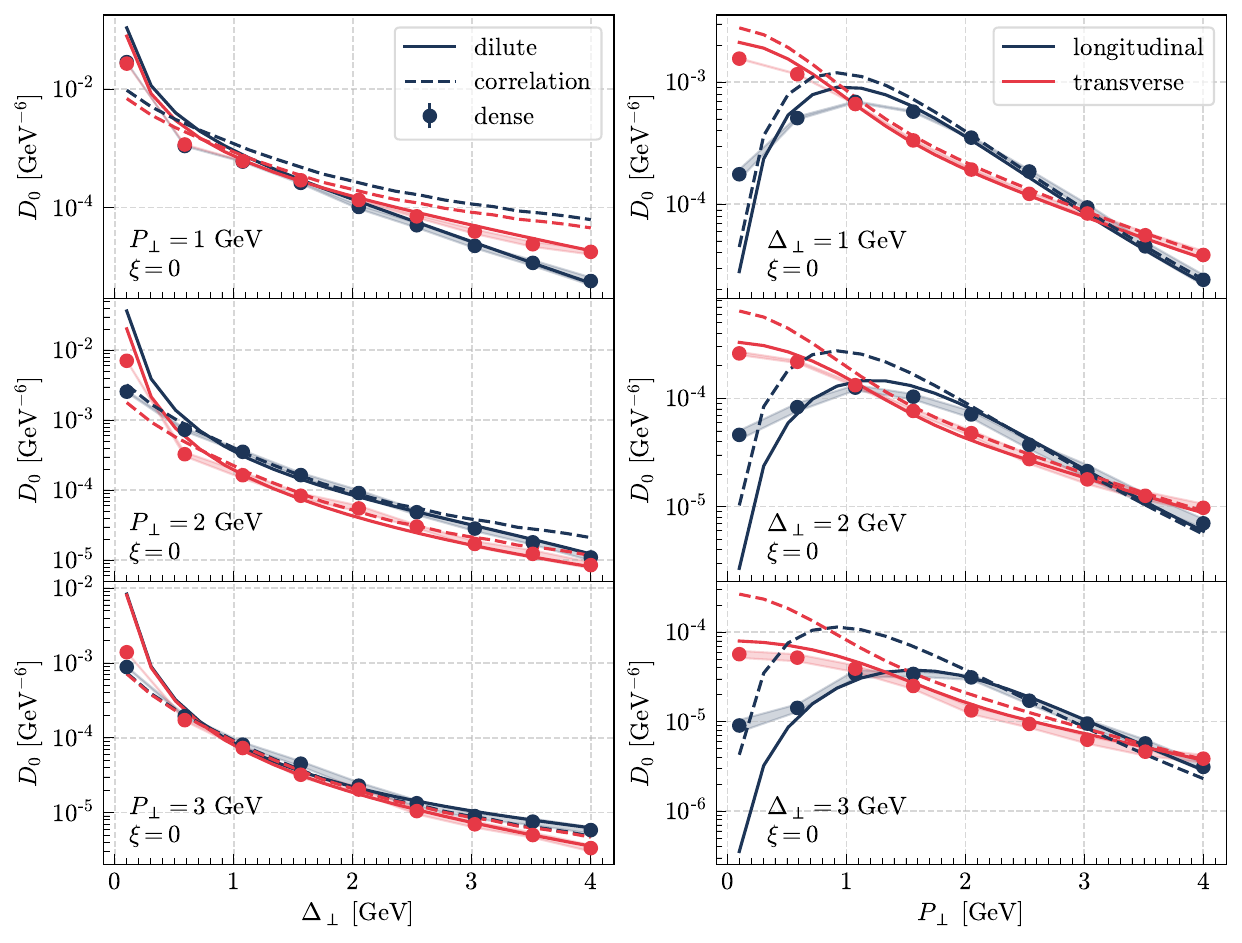}
		\caption{Left: Eikonal angle-averaged dijet cross section as a function of $\Delta_\perp$ and $P_\perp = 1,2,3$ GeV. Right: Eikonal angle-averaged dijet cross section as a function of $P_\perp$ and $\Delta_\perp = 1, 2, 3$ GeV. In all the figures we take $\xi=0$ and compare the dilute limit, represented by a continuous line, with the dense limit (dotted line) and its correlation limit (dashed line).}
		\label{fig:CGC_comp1}
	\end{figure}

	\Cref{fig:CGC_comp1} visually depicts the aforementioned comparison of the angle-averaged cross section in the dilute limit with the dense and correlation limits. In order to make a better comparison with the approach in~\cref{sec:DIS_dilute}, where the inclusion of a non-perturbative parameter was not required, we use a small value of $\Lambda_{\rm QCD} = 0.1$ GeV. As expected, all approaches converge in the limit $P_\perp \gg \Delta_\perp \gg Q_s$. Notably, the dilute limit exhibits a closer agreement with the dense limit than the correlation limit across a wider region of the kinematic space. This observation is particularly noteworthy considering the inherent analytical simplicity of the dilute limit expression. This finding suggests that the proposed approach offers a computationally efficient tool for calculations within a broader kinematic range compared to other limits.
	
\section{Conclusions}
\label{sec:conclusion}
	
	In this work we have started the study of the effects of non-eikonal corrections in the CGC on observables relevant for nuclear DIS, particularly at the energies of the future EIC. We have analyzed the effect of non-eikonal corrections on dijet production in DIS off nuclear targets. We have considered the next-to-eikonal corrections for this observable derived in~\cite{Altinoluk:2022jkk}. They require the knowledge of correlators of fields in the target beyond those computed in the MV model, specifically those between transverse and boost-enhanced (minus) components, and of the recoil ($x^-$-dependence) of the field. We have ignored the latter, while for the former we have developed
	%\textcolor{red}{an abelian}
	a linear model valid for large nuclei, which reproduces the MV model results for the correlator of boost-enhanced components and generalizes the one in~\cite{Cougoulic:2020tbc} going beyond spin exchange contributions.
	
	We have considered the unpolarized cross sections for dijet production in the approximation of a homogenous dilute nucleus, obtaining simple analytic expressions for the cross sections in the eikonal and next-to-eikonal approximations valid in the limit of total dijet momentum $P_\perp$  and dijet momentum imbalance $\Delta_\perp$ larger than the saturation scale $Q_s$, $P_\perp, \Delta_\perp \gtrsim Q_s$.
	
	We have performed a numerical study of the results applied to $e^-+\text{Au}$ collisions at top EIC energies. We have found $\mathcal{O}(10\%)$ effects in the cross sections, particularly at large $P_\perp$, thus making the inclusion of non-eikonal corrections necessary for precision studies at the EIC. We have analyzed the azimuthal asymmetries between $\mathbf{P}$ and $\mathbf{\Delta}$, $P_\perp = |{\bf P}|$, $\Delta_\perp = |{\bf \Delta}|$. We have found that non-eikonal corrections induce odd azimuthal harmonics for the symmetric situation (i.e., jets with equal momentum fractions from the virtual photon) where they are absent in the eikonal approximation, therefore offering a clean signature of non-eikonal effects.
	
	Finally, in the eikonal approximation we have compared the results of our analytic expansion valid in the dilute limit of the target, and the full CGC results in the MV model and their correlation limit~\cite{Dominguez:2011wm}. We have shown that our analytic expressions match the correlation limit ones in the region where both should be simultaneously valid, $P_\perp\gg \Delta_\perp \gtrsim Q_s$. We have also observed that our simple analytic formulae reproduce very well the full CGC results in its validity region $P_\perp, \Delta_\perp \gtrsim Q_s$.	
	
%	\textcolor{red}{\bf P.A.: I would perform the following modification to the last paragraph to make a more complete outlook.}
%	
%	\textcolor{red}{
%	As an outlook, we leave for the future the study of the effects of additional dependences (e.g., on $x^-$) of the correlators of target fields and of models for non-homogeneous targets, the inclusion of next-to-next-to-eikonal corrections, the analysis of the implications of non-eikonal corrections on the relation between target averages of Wilson lines and TMDs, and the interplay between next-to-leading and next-to-eikonal effects.}
	
	Given the focus of this paper solely on nuclear DIS, one crucial future direction is the extension to the case of DIS on light nuclei or protons. This extension necessitates the inclusion of longitudinal correlations and two-body operators, neglected in~\cref{sec:notation_LCWF}. Furthermore, when $P_\perp, \Delta_\perp < Q_s$ multi-gluon exchanges become more significant, requiring the application of resummation techniques. Consequently, exploring the dense limit becomes paramount for a comprehensive understanding of DIS at next-to-eikonal accuracy across the full kinematic regime.
	
For a complete picture of next-to-eikonal corrections, it would be interesting to incorporate the effects of the $x^-$ dependence of the field, as well as quark or antiquark exchanges with the background field. Additionally, incorporating next-to-next-to-eikonal corrections would enhance the calculation's precision, as these corrections include contributions from the transverse gradient of the $A^-$ field component as well as the $A^+$ field component. Finally, investigating the relationship between target averages of Wilson lines and TMD distributions, and the interplay between next-to-leading and next-to-eikonal effects, could offer valuable insights into non-eikonal effects.

\section*{Acknowledgments}
\label{sec:acknowledgements}
	
	We thank Guillaume Beuf, Florian Cougoulic and Fabio Dom\'{\i}nguez for useful discussions.
	P.A. and N.A. have received financial support from Xunta de Galicia (Centro singular de
investigaci\'on de Galicia accreditation 2019-2022, ref. ED421G-2019/05), by the European Union ERDF,
by the ``Mar\'{\i}a de Maeztu" Units of Excellence program MDM2016- 0692, and by the Spanish Research
State Agency under project PID2020-119632GBI00. This work was performed in the framework of the
European Research Council project ERC-2018-ADG-835105 YoctoLHC and the MSCA RISE 823947
``Heavy ion collisions: collectivity and precision in saturation physics" (HIEIC) and has received
funding from the European Union's Horizon 2020 research and innovation programme under grant
agreement No. 824093. P.A. has been supported by Conseller\'{\i}a de Cultura, Educaci\'on e Universidade
of Xunta de Galicia under the grant ED481B-2022-050.
	
\appendix
	
\section{Inclusive dijet production in the dense limit}
\label{sec:dense_limit}
	
	This section briefly reviews the evaluation of the differential cross section introduced in Section~\ref{sec:DIS_Neik} within the dense limit. Here, multiple gluon exchanges between the quark-antiquark pair and the dense nucleus requires resummation techniques. For a Gaussian (MV) correlation of gluon fields, this resummation can be performed analytically, resulting in the following nonlinear expressions, as derived in~\cite{Dominguez:2011wm}, for the dipole operator
	\begin{align}
	\label{eq:mvave}
		d({\bf v}, {\bf w}) = e^{-\Sigma({\bf v} - {\bf w})}, \quad
		\Sigma({\bf r}) = \frac{Q_s^2}{4} {\bf r}^2 \ln \left( \frac{1}{|{\bf r}| \Lambda_{\rm QCD}} + e \right)
	\end{align}
	and the quadrupole operator:
	\begin{align}
		Q({\bf w}', {\bf v}', {\bf v}, {\bf w}) &= d({\bf w}, {\bf v}) d({\bf w}', {\bf v}')
		\left[
		\left(
		\frac{F({\bf w}', {\bf v}, {\bf v}', {\bf w}) + \sqrt{\gamma}}{2\sqrt{\gamma}} - \frac{F({\bf w}', {\bf v}', {\bf v}, {\bf w})}{\sqrt{\gamma}}
		\right) e^{\frac{N_c}{4} \sqrt{\gamma}}\right.
		\nonumber \\ & 
		-
		\left.\left(
		\frac{F({\bf w}', {\bf v}, {\bf v}', {\bf w}) - \sqrt{\gamma}}{2\sqrt{\gamma}} - \frac{F({\bf w}', {\bf v}', {\bf v}, {\bf w})}{\sqrt{\gamma}}
		\right) e^{-\frac{N_c}{4} \sqrt{\gamma}} 
		\right]
		e^{-\frac{N_c}{4} F({\bf w}', {\bf v}, {\bf v}', {\bf w}) + \frac{1}{2 N_c} F({\bf w}', {\bf v}', {\bf v}, {\bf w})}.
	\end{align}
This expression involves the following auxiliary functions:
	\begin{align}
		F({\bf w}', {\bf v}', {\bf v}, {\bf w}) &= \frac{1}{C_F} \ln \frac{d({\bf w}' - {\bf v}) d({\bf v}' - {\bf w})}{d({\bf w}' - {\bf w})d({\bf v}' - {\bf v})}, \\
		\gamma({\bf w}', {\bf v}', {\bf v}, {\bf w}) &= F^2({\bf w}', {\bf v}, {\bf v}', {\bf w}) + \frac{4}{N_c^2} F({\bf w}', {\bf v}', {\bf v}, {\bf w})
		F({\bf w}', {\bf 2}, {\bf v}, {\bf v}'),
	\end{align}
	where $\gamma \equiv \gamma({\bf w}', {\bf v}', {\bf v}, {\bf w})$.
	
	This approach not only reproduces the dilute approximation from Section~\ref{sec:DIS_dilute} but also captures higher-order field correlators, extending its validity to lower momentum regimes. However, its application in this study was limited due to the current inability to resum non-eikonal field correlators.
	
	The differential cross section can be calculated using the aforementioned results in~\cref{eq:xsectionL_eik}. Furthermore, the Fourier harmonics can be efficiently computed using the following formula~\cite{Boussarie:2021ybe}:
	\begin{align}
		D_{n,\lambda}(P_\perp, \Delta_\perp) &= (-1)^n \int_{{\bf B}, {\bf r}, {\bf r}'} J_n(|{\bf B}| \Delta_\perp) J_n(|{\bf r}-{\bf r}'| P_\perp)
		\mathcal{C}_{\lambda}({\bf r},{\bf r}') \
		\Xi({\bf B} + z_2 {\bf r}, {\bf B} - z_1 {\bf r}, -z_1 {\bf r}', z_2 {\bf r}'),
	\end{align}
	where
	\begin{align}
		\Xi({\bf w}', {\bf v}', {\bf v}, {\bf w}) &= Q({\bf w}', {\bf v}', {\bf v}, {\bf w}) - d({\bf w}', {\bf v}') - d({\bf w}, {\bf v}) + 1.
	\end{align}
	
\subsection{The correlation limit}
\label{sec:correlation_limit}

	In the correlation limit, where $\Delta_\perp \ll P_\perp$ (or equivalently, $|{\bf r}|,|{\bf r}'|\ll |{\bf B}|$), we can expand the dipole and quadrupole operators in powers of the dipole sizes. This expansion allows to obtain analytical expressions for the differential cross sections in the eikonal approximation~\cite{Dominguez:2011wm}:
	\begin{align}
	\label{eq:aux16}
		&\frac{d \sigma^{\gamma^*_L + A \to q  \bar{q} + X}}{d^2 {\bf P} d^2 {\bf \Delta} d \eta_1  d \eta_2} \Bigg|_{\rm Eik.}
		=
		8 \alpha_{\rm em} e_f^2 Q^2 S_\perp \delta_z  z_1^3 z_2^3
		\frac{P_\perp^2}{(P_\perp^2 + \epsilon_f^2)^4} 
		\left[
		\alpha_s x G(\Delta_\perp) + \alpha_s x h(\Delta_\perp) \cos 2 \phi
		\right],
		\\
		\label{eq:aux17}
		&\frac{d \sigma^{\gamma^*_T + A \to q  \bar{q} + X}}{d^2 {\bf P} d^2 {\bf \Delta} d \eta_1  d \eta_2} \Bigg|_{\rm Eik.}
		=
		\alpha_{\rm em} e_f^2 S_\perp \delta_z  z_1 z_2 (z_1^2 + z_2^2)
		\frac{P_\perp^4 + \epsilon_f^4}{(P_\perp^2 + \epsilon_f^2)^4} 
		\left[
		\alpha_s x G(\Delta_\perp) 
		- \frac{2 P_\perp^2 \epsilon_f^2}{P_\perp^4 + \epsilon_f^4}
		\alpha_s x h(\Delta_\perp) \cos 2 \phi
		\right].
	\end{align}
	Here, $x G(\Delta_\perp)$ and $x h(\Delta_\perp)$ represent the Weizs\"acker-Williams gluon distribution and its linearly polarized partner, respectively. In the MV model and within the correlation limit, these functions read
	\begin{align}
		\label{eq:aux18}
		\alpha_s x G(\Delta_\perp)  &= \frac{N_c^2-1}{(2 \pi)^3 N_c} \int_0^\infty dB B J_0(\Delta_\perp B)
		\left[
		1 - e^{- \frac{2 N_c^2}{N_c^2 -1} \Sigma(B)}
		\right] 
		\frac{1}{\Sigma(B)} \left[
		\frac{d^2}{dB^2} + \frac{1}{B}\frac{d}{dB}
		\right]
		\Sigma(B),
		\\
		\label{eq:aux19}
		\alpha_s x h(\Delta_\perp)  &= -\frac{N_c^2-1}{(2 \pi)^3 N_c} \int_0^\infty dB B J_2(\Delta_\perp B)
		\left[
		1 - e^{- \frac{2 N_c^2}{N_c^2 -1} \Sigma(B)}
		\right] 
		\frac{1}{\Sigma(B)} \left[
		\frac{d^2}{dB^2} - \frac{1}{B}\frac{d}{dB}
		\right]
		\Sigma(B).
	\end{align}
	
	The expressions presented here hold validity only in the correlation limit, $P_\perp \gg \Delta_\perp$. Conversely, the expressions derived in~\cref{eq:xSectionL_Eik,eq:xSectionT_Eik} are valid within the dilute limit, where both $P_\perp$ and $\Delta_\perp$ are significantly larger than the saturation scale $Q_s$. Therefore, we anticipate substantial overlap between the correlation and dilute limits in the region where $P_\perp \gg \Delta_\perp \gtrsim Q_s$.
	
\begin{figure}[h!]
		\centering
		\includegraphics[scale=0.8]{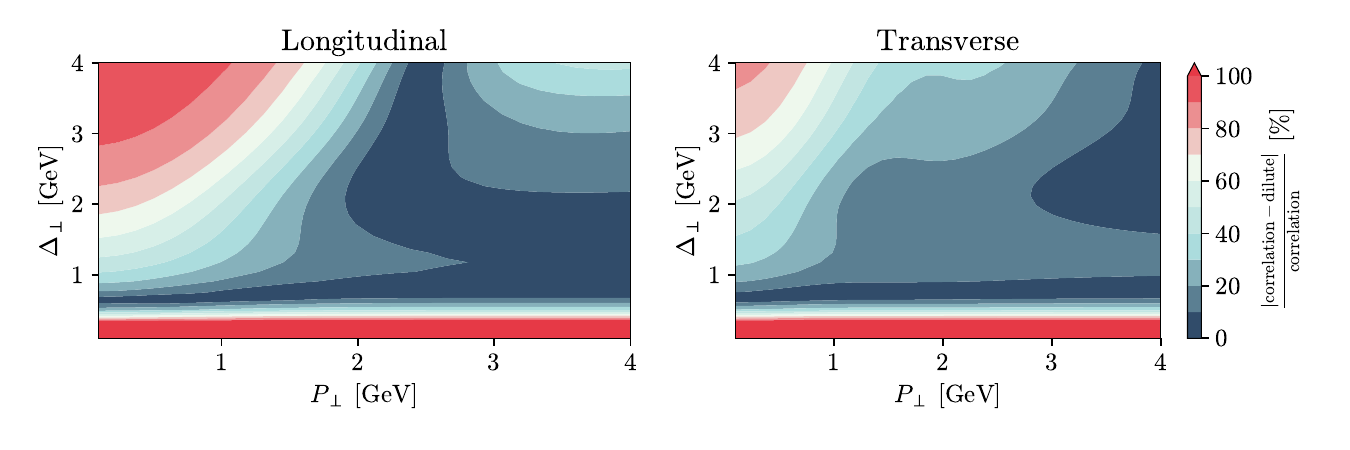}
		\caption{Relative difference between the dilute and correlation limits of the longitudinal (left) and transverse (right) dijet cross sections in the CGC as a function of $P_\perp$ and $\Delta_\perp$.}
		\label{fig:TMD_comparison}
	\end{figure}

	Indeed,~\cref{fig:TMD_comparison} visually depicts the relative difference between the dilute and correlation limits as a function of $P_\perp$ and $\Delta_\perp$. For better comparison, we employ a smaller value of $\Lambda_{\rm QCD} = 0.1$ GeV. As the figure demonstrates, good agreement between both approaches is observed within the expected kinematic regime of $P_\perp \gg \Delta_\perp \gtrsim Q_s$.
	
	Finally,~\cref{eq:xSectionL_Eik,eq:xSectionT_Eik} can be reduced to the following expressions in the correlation limit ($P_\perp \gg \Delta_\perp$):
	\begin{align}
		\label{eq:aux20}
		\frac{d \sigma^{\gamma^*_L + A \to q  \bar{q} + X}}{d^2 {\bf P} d^2 {\bf \Delta} d \eta_1  d \eta_2} \Bigg|_{\rm Eik.}
		&= \frac{2 C_L}{\Delta_\perp^2} \frac{Q^2 P_\perp^2}{(P_\perp^2 + \epsilon_f^2)^4}(1 + \cos 2 \phi),
		\\
		\label{eq:aux21}
		\frac{d \sigma^{\gamma^*_T + A \to q  \bar{q} + X}}{d^2 {\bf k}_1 d \eta_1 d^2 {\bf k}_2 d \eta_2} \Bigg|_{\rm Eik.}
		&= \frac{C_T}{\Delta_\perp^2} \frac{P_\perp^4 + \epsilon_f^4}{(P_\perp^2 + \epsilon_f^2)^4} \Bigg[ 1 - \frac{2 P_\perp \epsilon_f}{P_\perp^4 + \epsilon_f^4} \cos 2 \phi \Bigg].
	\end{align}
	
	Using the saturation momentum definition in~\cref{eq:saturation_momentum} and comparing~\cref{eq:aux16,eq:aux17} with~\cref{eq:aux20,eq:aux21}, we can analytically derive the Weizs\"acker-Williams gluon distributions in the limit of $\Delta_\perp \gg Q_s$:
	\begin{align}
		x G(\Delta_\perp) =  x h(\Delta_\perp) = \frac{2 N_c C_F \tilde{\mu}^2 S_\perp}{\pi^2 \Delta_\perp^2}.
	\end{align}
	Interestingly, this result aligns perfectly with the one obtained from perturbative QCD~\cite{Dominguez:2011wm}.

\bibliography{mybib}

\end{document}